\documentclass[english,sort&compress]{iopart}
\usepackage[T1]{fontenc}
\usepackage[latin9]{inputenc}
\usepackage{amsbsy}
\usepackage{amstext}
\usepackage{geometry}
\usepackage[numbers]{natbib}

\makeatletter
\usepackage{iopams}
\usepackage{setstack}

\usepackage{hyperref}

\newcommand{\eqref}[1]{(\ref{#1})}

\makeatother

\usepackage{babel}
\begin{document}
\title[The Algebra of the Pseudo-Observables I]{The Algebra of the Pseudo-Observables I: Why Quantum Mechanics is
the ultimate description of Reality}
\author{Edoardo Piparo\footnote{A.I.F. Associazione per l'Insegnamento della Fisica - Gruppo Storia della Fisica: \url{http://www.lfns.it/STORIA/index.php/it/chi-siamo}}}
\address{{\large I.I.S. ``Elsa Morante'', Viale Francesco Selmi 16, I-41049
Sassuolo (MO), Italy}}
\ead{{\large edoardo.piparo@istruzione.it}}
\begin{abstract}
This paper is the first of several parts introducing a new powerful
algebra: the algebra of the pseudo-observables. This is a C{*}-algebra
whose set is formed by formal expressions involving observables. The
algebra is constructed by applying the Occam's razor principle, in
order to obtain the minimal description of physical reality.

Proceeding in such a manner, every aspect of quantum mechanics acquires
a clear physical interpretation or a logical explanation, providing,
for instance, in a natural way the reason for the structure of complex
algebra and the matrix structure of the formulation of Werner Heisenberg
of quantum mechanics.

Last but not least, the very general hypotheses assumed, allow one
to state that quantum mechanics is the unique minimal description
of physical reality. 
\end{abstract}

\maketitle
\global\long\def\ket#1{|#1\rangle}%

\global\long\def\bra#1{\langle#1|}%

\global\long\def\inprod#1#2{\left\langle #1\mid#2\right\rangle }%

\global\long\def\ideq{:=}%

\global\long\def\imp#1{#1_{\mathrm{I}}}%

\global\long\def\rp#1{#1_{\mathrm{R}}}%

\section{Introduction}

\subsection{A brief historical background}

After a long and hard working, started from the formula proposed by
Planck in 1900 for the black-body radiation and Einstein's ideas on
light quanta and those of Bohr on the constitution of atoms, quantum
mechanics suddenly arises between 1925 and 1927, thanks to some brilliant
scientists such as Heisenberg, Born, Jordan, Schrödinger, Dirac, Pauli.
The new formalism has quickly found a resounding success in justifying
and predicting the phenomena of the atomic world, but it also immediately
proves itself much more abstract than that of classical physics, and
therefore more difficult to be acquired.

A few months after their publication in 1926, the four Schrödinger's
works\citep{Schroedinger1926,Schroedinger1926a,Schroedinger1926b,Schroedinger1926c}
on quantization as an eigenvalue problem, collected in one volume\citep{Schroedinger1927}
together with others of the same author, formed the first manual of
wave mechanics. It was, however, only a manual for a few specialists,
soon followed by new ones, among which outstanding is the fundamental
text of Dirac\citep{Dirac1930}, written so definitive that the first
edition of 1930 has remained virtually unchanged for ten of its twelve
chapters until the fourth and final version of 1958 and reprinted
for the seventh time in 1976. Numerous presentations of quantum mechanics,
the most comprehensive and educationally oriented, have followed in
the now long period of time that separates us from those glorious
years, in which the intellectual liveliness and mathematics knowledge
of some young researchers, together with the genius and the experience
of more mature scientists have produced one of the greatest revolutions
in the history of thought.

The reorientation of perspective introduced by quantum mechanics in
the way of thinking the natural phenomena affects not only the attitude
of people of science, but in the intimate touches the mentality of
the whole mankind. Accustomed to nineteenth-century physics - the
so-called classical physics based on the mechanics of Galileo and
Newton - one could conceive of the entire universe in mechanistic
evolution, according to a deterministic cause and effect chain of
relations. Emblematic of this is the reductionist position of Laplace,
that is unexpectedly denied by the conclusions reached by the physicists
in building a theory able to account for the new experimental data
accumulated at the end of the Nineteenth and in the first quarter
of the Twentieth. Heisenberg discovery of limitations on observation,
which prevent one to determine exactly the initial conditions from
which depend the time evolution of classical physics, puts on evidence
the participation of the observer itself to the construction of the
phenomenon and it is a stimulus for philosophical insights which resized
the human-nature relationship (the first part of this short historical
survey is partly based on the preface of the Boffi's book\citep{Boffi2010}).

Despite the successes of quantum mechanics in explaining physical
phenomena, the \emph{interpretation of quantum mechanics}, however,
was a much more controversial issue, that may be regarded also now
as an opened one. An \emph{interpretation of quantum mechanics} is
a set of statements which attempts to explain how quantum mechanics
informs our understanding of nature. Although quantum mechanics has
held up to rigorous and thorough experimental testing, many of these
experiments are open to different interpretations.

A particularly problematic aspect of the matter is that it seems to
be more in the fields of interest of philosophers of physics than
of those of physicists themselves. The latter however felt and feel
the need for an interpretation of the \emph{mathematical formalism}
of quantum mechanics, specifying the\emph{ physical meaning} of the
mathematical entities of the theory.

However the issue is so slippery that even Dirac, in the new edition
of 1958 of his fundamental treatment\citep{Dirac1930}, states: «the
main object of physical science is not the provision of pictures,
but is the formulation of laws to the discovery of new phenomena»,
that sounds like a capitulation declaration.

Nevertheless, several researchers fronted the issue, giving rise to
a number of contending schools of thought, differing over whether
quantum mechanics can be understood to be deterministic, which elements
of quantum mechanics can be considered ``real'' and other matters.

The definition of quantum theorists' terms, such as wavefunctions
and matrix mechanics, progressed through many stages. For instance,
Erwin Schrödinger originally viewed the electron's wavefunction as
its charge density smeared across the field, whereas Max Born\citep{Born1926}
reinterpreted it as the electron's probability density distributed
across the field.

Among the plethora of interpretations, I limit myself here to recall
only a few of them.

The \emph{Copenhagen interpretation} is the ``standard'' interpretation
of quantum mechanics formulated by Niels Bohr and Werner Heisenberg
while collaborating in Copenhagen around 1927. According to John G.
Cramer, ``Despite an extensive literature which refers to, discusses,
and criticizes the Copenhagen interpretation of quantum mechanics,
nowhere does there seem to be any concise statement which defines
the full Copenhagen interpretation.''\citep{Cramer1986}

The \emph{many-worlds interpretation} is an interpretation of quantum
mechanics in which the phenomena associated with measurement are claimed
to be explained by \emph{decoherence}, which occurs when states interact
with the environment producing entanglement, repeatedly splitting
the universe into mutually unobservable alternative histories-distinct
universes within a greater \emph{multiverse}. The original formulation
is due to Hugh Everett in 1957\citep{Everett1957}. Later, this interpretation
was popularized and renamed \emph{many-worlds} by Bryce Seligman DeWitt
in the 1960s and 1970s\citep{DeWitt1970}.

The \emph{consistent histories} approach is intended to give a modern
interpretation of quantum mechanics, generalizing the conventional
Copenhagen interpretation\citep{Griffiths1984}. This interpretation
of quantum mechanics is based on a consistency criterion that then
allows probabilities to be assigned to various alternative histories
of a system such that the probabilities for each history obey the
rules of classical probability while being consistent with the Schrödinger
equation. In contrast to some interpretations of quantum mechanics,
particularly the Copenhagen interpretation, the framework does not
include ``wavefunction collapse'' as a relevant description of any
physical process, and emphasizes that measurement theory is not a
fundamental ingredient of quantum mechanics.

The \emph{relational quantum mechanics} is an interpretation of quantum
mechanics which treats the state of a quantum system as being observer-dependent,
that is, the state is the relation between the observer and the system.
This interpretation was first delineated by Carlo Rovelli\citep{Rovelli1996},
and has since been expanded upon by a number of theorists. The essential
idea behind it, inspired by special relativity, is that different
observers may give different accounts of the same series of events:
for example, to one observer at a given point in time, a system may
be in a single, ``collapsed'' eigenstate, while to another observer
at the same time, it may be in a superposition of two or more states.
Consequently, if quantum mechanics is to be a complete theory, relational
quantum mechanics argues that the notion of ``state'' describes
not the observed system itself, but the relationship, or correlation,
between the system and its observer(s). The state vector of conventional
quantum mechanics becomes a description of the correlation of some
degrees of freedom in the observer - intended here as a generic physical
object, whether or not conscious or macroscopic - with respect to
the observed system. Any ``measurement event'' is seen simply as
an ordinary physical interaction, an establishment of the above sort
of correlation. Thus the physical content of the theory has to do
not with objects themselves, but the relations between them.

The \emph{objective collapse }theories differ from the Copenhagen
interpretation in regarding both the wavefunction and the process
of collapse as ontologically objective. The most well-known examples
of such theories are: the \emph{Ghirardi-Rimini-Weber theory} - first
reported in 1985\citep{G.C.Ghirardi1985} - in which, as an attempt
to avoid the measurement problem in quantum mechanics, is proposed
that wave function collapse happens spontaneously; and the \emph{Penrose
interpretation}, in which it is proposed that a quantum state remains
in superposition until the difference of space-time curvature attains
a significant level\citep{Penrose1989}.

\subsection{The quest for a new formulation}

The proliferation of interpretations proposed is a clear symptom of
the fact that, in contrast to what stated by Dirac, the laws which
rule the phenomena are not sufficient to provide a complete understanding
of Nature. For this to happen, such laws must not contain any \textquotedblleft gray
area\textquotedblright , i.e. shall not to depend on entities or processes
whose nature is not clear. But the traditional formulation of quantum
mechanics actually depends on these. Wavefunctions, or, more generally
speaking, vector-states, haven't a clear and complete physical interpretation,
and it's not even clear if they are or aren't ontologically real.
The measurement process is account only for his final effect on vector
states, without any reference to its time-development, and is not
describable as a physical process. These issues are related to the
choose of the fundamental entities and of the fundamental laws of
quantum mechanics, i.e., ultimately, to its mathematical formulation.

The only possibility to resolve this is therefore to look for a new
formulation of quantum mechanics.

The earliest versions of quantum mechanics were formulated in the
first decade of the 20th century, as theories of matter and electromagnetic
radiation, in order to front the problems arisen when the atomic theory
and the corpuscular theory of light first came to be widely accepted
as scientific facts.

Early quantum theory was significantly reformulated in 1925-1926 following
a dual path. Werner Heisenberg\citep{Heisenberg1925}, Max Born and
Pascual Jordan\citep{M.Born1925,M.Born1925a} gave rise to \emph{matrix
mechanics}; Louis de Broglie\citep{Broglie1925} and Erwin Schrödinger\citep{Schroedinger1926d}
created \emph{wave mechanics}.

Although Schrödinger himself after a year proved the equivalence of
his wave-mechanics and Heisenberg's matrix mechanics, the reconciliation
of the two approaches and their modern abstraction as motions in Hilbert
space is generally attributed to Paul Dirac, who wrote a lucid account
in his 1930 classic \emph{The Principles of Quantum Mechanics}\citep{Dirac1930}.
Dirac introduced the \emph{bra-ket} notation, together with an abstract
formulation in terms of the \emph{Hilbert space} used in functional
analysis; he showed that Schrödinger's and Heisenberg's approaches
were two different representations of the same theory, and found a
third, more general one, which represented the dynamics of the system.

The first complete mathematical formulation of this approach, known
as the \emph{Dirac\textendash von Neumann axioms}, is generally credited
to John von Neumann's 1932 book \emph{Mathematical Foundations of
Quantum Mechanics}\citep{Neumann1932}, although Hermann Weyl had
already referred to Hilbert spaces (which he called \emph{unitary
spaces}) in his 1927 classic paper\citep{Weyl1927} and book\citep{Weyl1928}.

A new formulation of quantum mechanics had to wait until 1948, when
Richard Feynman, recovering an idea presented by Dirac in his 1933
paper\citep{Dirac1933}, introduced his \emph{path integral formulation}\citep{Feynman1948},
in which a quantum-mechanical amplitude is considered as a sum over
all possible classical and non-classical paths between the initial
and final states. This is the quantum-mechanical counterpart of the
\emph{action principle} in classical mechanics and allows a quantum
mechanical formulation based on Lagrangian rather than Hamiltonian,
which is more desirable in certain theoretical contexts.

All of these formulation, however, rely on the same axiomatic context
and share the ``gray areas'' proper of the \emph{Copenhagen interpretation},
that is their common framework.

The issues arisen by the von Neumann's postulate on the collapse of
the wavefunction, however, stimulated the development, in the intervening
70 years, of new formulations of quantum mechanics in which the overcoming
of the \emph{problem of measurement} was central. \emph{Many-worlds
}interpretation, \emph{consistent histories} approach, \emph{objective
collapse }theories and \emph{quantum logic} arose in such a context.
\emph{Quantum logic} is a set of rules for reasoning about propositions
which takes the principles of quantum theory into account. This research
area and its name originated in the 1936 paper by Garrett Birkhoff
and John von Neumann\citep{Birkhoff1936}, who attempted to reconcile
the apparent inconsistency of classical logic with the facts concerning
the measurement of complementary variables in quantum mechanics, such
as position and momentum.

On another side, some of the originators of quantum theory (notably
Einstein and Schrödinger) were unhappy with what they thought were
the philosophical implications of quantum mechanics. In particular,
Einstein took the position that quantum mechanics must be incomplete,
which motivated research into so-called \emph{hidden-variable} theories.
In his 1926 paper\citep{Born1926}, Max Born, was the first to clearly
enunciate the probabilistic interpretation of the quantum wavefunction,
which had been introduced by Erwin Schrödinger earlier in the same
year. Born's interpretation of the wavefunction was criticized by
Schrödinger, who had previously attempted to interpret it in real
physical terms, but Albert Einstein's response\citep{Einstein1926},
contained in a letter sent to Max Born in 1926, became one of the
earliest and most famous assertions that quantum mechanics is incomplete:
\begin{quotation}
Quantum mechanics is certainly imposing. But an inner voice tells
me that this is not yet the real thing. The theory says a lot, but
does not really bring us any closer to the secret of the 'old one'.
I, in any rate, am convinced that \emph{He} is not playing at dice.
\end{quotation}
Shortly after making his famous comment, Einstein attempted to formulate
a deterministic counterproposal to quantum mechanics. On 5 May 1927
he read a paper to the Prussian Academy of Sciences in Berlin entitled:
``\emph{Bestimmt Schrödinger's Wellenmechanik die Bewegung eines
Systems vollständig oder nur im Sinne der Statistik?}'' (\textquotedblleft \emph{Does
Schrödinger\textquoteright s wave mechanics determine the motion of
a system completely or only in the statistical sense?}\textquotedblright );
but, just a few weeks later, as the paper was being prepared for publication
in the academy's journal, he decided to withdraw it, possibly because
he discovered that implied non-separability of entangled systems could
not be eliminated, as he had hoped\citep{Baggott2011}.

At the Fifth Solvay Congress, held in Belgium in October 1927, Louis
de Broglie presented his own version of a deterministic hidden-variable
theory\citep{Broglie1928}, apparently unaware of Einstein's aborted
attempt earlier in the year. In his theory, every particle had an
associated, hidden ``\emph{pilot wave}'' which served to guide its
trajectory through space. de Broglie generalized the pilot-wave dynamics
to many-body system, presenting also some elementary applications
(interference, scattering). The theory was discussed extensively,
with reactions, comments and critics of Pauli, Born, Brillouin, Einstein,
Kramer, Lorentz, Schrödinger and others; but de Broglie defended it
fairly well\citep{Bacciagaluppi2013}. However he abandoned the theory
shortly thereafter.

The general interest on this approach was lost when von Neumann, in
his 1932 book\citep{Neumann1932}, claimed to prove the impossibility
of theories which, by using the so called hidden variables, attempt
to give a deterministic explanation of quantum mechanical behaviors.

It's worth recalling that in 1935 Grete Hermann criticized the von
Neumann proof on a fundamental point\citep{Hermann1935}. Her work,
however, remained substantially unnoticed by the physics community
for several decades.

In the meantime, in 1952, David Bohm, dissatisfied with the prevailing
orthodoxy, developed a theory\citep{Bohm1952,Bohm1952a} that is essentially
the same as de Broglie's pilot wave theory, the only difference being
that de Broglie's dynamics is formulated in terms of velocity rather
than acceleration. However the general interest in the theory was
little, till when, in 1966, John Bell\citep{Bell1966} rediscovered
the results of Grete Hermann's work, i.e. that ``von Neumann's no-hidden-variables
proof was based on an assumption that can only be described as silly''\citep{Mermin1993}.
The von Neumann argumentation, in fact, demonstrates only that hidden-variables
theories\emph{ }must be \emph{nonlocal}.

Unlike Hermann's one, Bell's critique had a great foundational impact
and Bell himself, during the sixties, the seventies and the eighties,
became the principal proponent of the now called \emph{de Broglie-Bohm
theory}, to which his 1987 book\citep{Bell1987} contains an yet unsurpassed
introduction.

The theory, anyway, remains still controversial. In August 2011, Roger
Colbeck and Renato Renner proved\citep{Colbeck2011} that, under the
assumption that measurements can be chosen freely, no extension of
quantum theory, whether using hidden variables or otherwise, can give
more information about the outcomes of future measurements than quantum
theory itself. In January 2013, however, Giancarlo Ghirardi and Raffaele
Romano described a model which, ``under a different free choice assumption
{[}\ldots{]} violates {[}the statement by Colbeck and Renner{]} for
almost all states of a bipartite two-level system, in a possibly experimentally
testable way''\citep{Ghirardi2013}.

Moreover, if on one side, it seems to resolve the measurement problem\citep{Maudlin1995},
on the other it still relies on the framework of the Dirac-von Neumann
axioms and on the concept of wavefunction, which is also regarded
as ontologically real.

Another interesting line of research in the field of quantum mechanics
formulations is that of \emph{algebraic} theories\citep{Landsman2009}.
 Algebraic quantum mechanics is an abstraction and generalization
of the Hilbert space formulation. In the development of the algebraic
approach a major role was played by von Neumann. His joint paper with
Jordan and Wigner\citep{Jordan1934} was, in fact, one of the first
attempts to go beyond Hilbert space of state vectors. Subsequently,
von Neumann developed the mathematical theory of operator algebras
in a series of outstanding papers\citep{Murray1936,Murray1937,Neumann1940,Murray1943,Neumann1949}.
Notwithstanding his efforts, von Neumann's attempts to apply this
theory to quantum mechanics were unsuccessfully\citep{Redei1996}.
The operator algebras that he introduced, now called \emph{von Neumann
algebras}, however, still play a central role in the algebraic approach
to quantum theory.

In 1943, Gelfand and Naimark\citep{Gelfand1943} introduced the notion
of \emph{C{*}-algebras} (the term ``C{*}-algebra'' was, however,
introduced by Irving Segal in 1947\citep{Segal1947}), a generalization
of von Neumann ones, freeing themselves from the need to do reference
to operators on a Hilbert space. The framework of a C{*}-algebra appears
as an ideal one to find a new formulation to quantum mechanics and
is already nowadays largely utilized to represent the actual one based
on the Dirac-von Neumann axioms.

An original solution to the problem of the nature of wavefunctions
is to try to reformulate quantum mechanics in a way such that these
entities are no more fundamentals ones, but only mathematical instruments.
A significant upgrade in this direction was made with what that its
creators called ``\emph{QBism}'' - a particular form of \emph{Quantum
Bayesianism}, i.e. a formulation that uses a Bayesian approach to
the probabilities that appear in quantum theory\citep{Fuchs2011}.
According to QBism, wavefunctions are solely a mathematical tool that
an observer uses to assign his personal belief that a quantum system
will have a specific property. In this conception, a wavefunction
isn't ontologically real but merely reflects an individual's subjective
mental state. The proponents of QBism embrace the notion that until
an experiment is performed, \emph{its outcome simply does not exist}\citep{Fuchs2014}.

One of the criticisms on QBism is that it is unable to explain complex
macroscopic phenomena in terms of more primitive microscopic ones,
in the way that conventional quantum mechanics does. In order to obtain
this, QBism needs again a reformulation of quantum mechanics, based
on a new set of axioms.

\subsection{What it has been done in this paper}

It would have to been clear, at this point, how it is important to
find a new formulation of quantum mechanics, based on a new set of
axioms and on a new interpretation. At this regards, I agree both
with Rovelli\citep{Rovelli1996} and Fuchs thesis\citep{Fuchs2010}.
The point is that quantum mechanics was formulated making reference
to the notions of ``observer'' and ``measurement'' but without
a real analysis about them, that were, as a matter of facts, taken
as primitive. In order to overcome this weakness, it arose the idea
to remove the ``observer'' from the theory as quickly as possible
and, to do this, the general strategy was to reify or objectify all
the mathematical symbols of the theory, disregarding that this would
have implied an unclear physical interpretation of them.

In this paper I propose a really new formulation of quantum mechanics,
that, in my opinion, is free of ``gray areas'' and fully physically
transparent in its axioms.

The new formulation is carried out as an algebraic theory and turns
to be expressed in terms of a C{*}-algebra.

The starting point was a deep analysis of fundamental concepts, as
``physical reality'', ``observers'', ``physical quantities''
and ``measurements''. On the basis of this analysis and the assumptions
therein implied, I constructed an algebraic structure for ``observables'',
that turned to be a C{*}-algebra, making use essentially only of the
\emph{Occam's razor principle}, that in this context has to be meant
as follows: \emph{in building the formulation of a physical theory
it is necessary to avoid, as much as possible, the introduction of
entities not reducible to measurable physical properties}. The aim
is to obtain the minimal description of physical reality; whereas
the introduction of unnecessary unobservable entities leads to descriptions
depending on a potentially enormous number of arbitrary parameters.

Proceeding in such a manner, every aspect of quantum mechanics acquires
a clear physical interpretation or a logical explanation, providing,
for instance, in a natural way the reason for the structure of complex
algebra and the matrix structure of the formulation of Werner Heisenberg
of quantum mechanics.

Last but not least, the very general hypotheses assumed, allow one
to state that \emph{quantum mechanics is the unique minimal description
of physical reality}. 

\section{Spaces of Observables}

\subsection{Reality and observers\label{subsec:Realta-ed-osservatori}}

For a big part, the human doing is based on the assumption of the
existence of an ``objective reality''. Such an hypothesis may considered
the most fundamental \emph{archetype} of our minds, upon which depend
also the fundamental concepts of \emph{space}, meant as the ``container''
of reality elements, and of \emph{time}, meant as the ordered of evolution
of reality elements.

The objectivity of reality implies the total independence of the external
world from the observer. This assumption of independence is, however,
largely unfounded.

The knowledge of reality, indeed, is based on subjective experiences.
The «objectivity» dimension rises only by knowledge mediated by language.
A language, however, is based on a shared set of \emph{symbols} species
specific, subject, moreover, to variations as time goes on. Such a
kind of «objectivity» is, therefore, \emph{conventional} and strongly
tied to the observers.

In the last analysis, we can convincingly state that «reality» is
nothing more than a \emph{coherent and shared reconstruction of a
complex of subjective experiences}. The objectivity of reality, therefore,
doesn't transcend the subjective level, but merely implies to take
into account only a convenient part of it.

Ultimately, then, the so-called ``physical reality'' is not distinguishable
from the mental representation that a group of coherent observers
make of it. In such a context, the ``\emph{laws of Physics}'' reflect
rules and logical relations rather more proper of the \emph{observer}
than of the \emph{observed}. Then, therefore, I think that one can
only agree with the Bohr's quotation\citep{Bohr_Quote}:
\begin{quotation}
It is wrong to think that the task of physics is to find out how nature
\emph{is}. Physics concerns what we can say about nature
\end{quotation}
and with what states Fuchs in\citep{Fuchs_Quote}:
\begin{quotation}
With every quantum measurement set by an experimenter\textquoteright s
free will, the world is shaped just a little as it participates in
a kind of moment of birth.
\end{quotation}
So, I am convinced that by looking at the Universe the Human Being
can see the reflex of his most intimate nature. In fact, I believe
in the truthiness of Weyl's words\citep{Weyl1934}:
\begin{quotation}
The world exists only as met with by an ego, as one appearing to a
consciousness; the consciousness in this function does not belong
to the world, but stands out against the being as the sphere of vision,
of meaning, of image, or however else one may call it.
\end{quotation}

\subsection{Physical quantities and observables}

The aim of Physics is the study of the evolution of a portion of the
world (the \emph{physical system}) that surrounds us, characterized
by a certain number of measurable properties, whatever is the system
\emph{state} (on the concept of state I will return in subsection\emph{
\ref{subsec:Osservabili}}), called, as usual in quantum mechanics,
\emph{observables}. Observables the measurements of which results
in measures belonging to the same set, and therefore comparable, will
be of the same \emph{type}.

It is easily verified that this is an\emph{ equivalence relation}.
It is, then, reasonable to identify a \emph{physical quantity} with
an \emph{equivalence class} formed by observables of the same type.
A physical quantity is, therefore, a measurable property relative
to a set of physical systems, like, for instance, the distance between
two points, whereas the term observable will be reserved to the same
property relative to a \emph{specific} physical system.

By these definitions, it is possible to state that a physical quantity
is characterized by the set of the possible outcomes of an its measurement.
Since this set is the result of a series of measurement processes,
it will contain a finite number of possible outcomes. This number
can, eventually, tend to infinite, in the ideal limit of measurement
instruments whose resolution capability tends to zero or whose range
tends to infinite. These are, however, only \emph{mathematical} limits,
unattainable for every observer. As already pointed out, the problem
is that in the Dirac formulation of quantum mechanics there is an
implied \emph{platonic} vision on the nature of the world, with the
implicit assumption that it must exist a sort of ``world of ideas'',
in the form of an abstract mathematical model, disregarding the possibility
of an effective measurability or interpretability of the entities
therein introduced.

It is worth, therefore, to follow a different way, starting from an
accurate \textbf{physical} analysis of physical quantities and observables,
in order to try to build a fully \textbf{physically} interpretable
conceptual picture.

\subsection{Observables as stochastic variables\label{subsec:Osservabili}}

Imagine one having to measure a certain observable relative to a given
system. Due to issues related to the operating limits of the measuring
instrument, to fluctuations in the environmental conditions and in
the system itself, a single measurement won't be sufficiently reliable.

In ideal conditions, what one could do is to consider a set (\emph{ensemble})
of faithful copies of the system under examination, all in the same
conditions, i.e. in the same \emph{state}, and to perform the measurement
simultaneously in all the copies. Obviously the outcomes will be,
generally, different and therefore, in the so outlined measurement
process, the observable assumes the connotation of a \emph{stochastic
variable}.

In what follows, observables will be usually labeled with uppercase
letters, for instance $X$, whereas their possible outcomes will be
denoted with the corresponding lowercase letter, in our example $x$.
The set of all possible outcomes of the measurement of a given observable
will be called the\emph{ observable spectrum}.

Particularly important are the\emph{ constant observables}, that are
observables for which there is only a single possible measurement
outcome $c$. These observables will be labeled with the name or the
value of their constant outcome, that will be briefly called \emph{constant's
value}. Among them, remarkable examples are the \emph{null observable},
whose measurement outcome is always 0, and the \emph{unit observable},
whose measurement outcome is always 1.

Two observables will be called \emph{compatible} if it is possible
to measure both simultaneously. Evidently, each observable is compatible
with itself and a constant observable is compatible with every observable.

It is, finally, important to point out that an observable relative
to a certain time instant $t$ isn't generally compatible with the
same observable evolved to a different time instant $t'$, since they
cannot be measured \textbf{simultaneously} and\textbf{ independently}. 

\subsection{Spaces of compatible observables\label{subsec:Spazi-di-Oss-Comp}}

Given two compatible observables, $A$ and $B$, it is easy to define
their sum or product: it suffices to consider an observable which
has as possible outcomes of measuring, respectively, the sum and the
product of the possible outcomes of measurement of the two observables.
One can similarly define the difference of two observables, and their
quotient, provided, however, for the latter, that the observable in
the denominator \emph{doesn't admit $0$ as a possible outcome of
measurement}. The observables obtained with these operations are clearly
compatible with starting ones.

According to the definition, the sum of two compatible observables
satisfies the commutative property, the associative one, the existence
of a neutral element (the null observable) and opposite element (the
observable whose possible outcomes of measurement are the opposites
of the one considered). For the product are valid the commutative
property, the associative one and the existence of neutral element
(the unit observable). The multiplication also distributes over addition.
Not all observables different from the null one, however, admit the
inverse element (those do not admit it are non-constant observables
that contain the zero in their spectrum).

The set of observables compatible with a given observable $A$, equipped
with the operations of sum and product thus defined, has, therefore,
the algebraic structure of a \emph{commutative unitary ring}. This
ring is not, however, an \emph{integral domain}, since, in general,
the \emph{zero-product property} isn't valid.

If we, then, identify constant observables with their values, it is
easy to see that every set consisting of mutually compatible observables,
provided with the operations of addition and multiplication by a constant
coefficient and closed with respect to these operations, has the structure
of a \emph{vector space} over the field $\mathbb{R}$ of the real
numbers. This will be called a \emph{space of compatible observables}.

It is also possible to introduce an equivalence criterion between
observables: \emph{two observables are equivalent if their difference
is the null observable}. It is easily verified that the\emph{ reflexive},
\emph{symmetric} and \emph{transitive} properties hold for the so-defined
criterion, as required for an equivalence relation.

Equivalent observables are physically indistinguishable and will be,
therefore, identified as the \emph{sam}e observable.

It is, finally, possible to introduce a \emph{partial order relation}
between two compatible observables. We will say that between two compatible
observables of the same type $A$ and $B$ it holds the relation $A\leq B$,
if for every simultaneous measurement of the two observable, of outcomes
respectively $a$ and $b$, one has always $a\leq b$. 

\subsection{Projectors}

In the theory of observables that I am outlining, of a particular
importance are the \emph{projectors} (the name is given in analogy
to that of the projection operators of the Dirac formulation, to which
they correspond). A projector is an observable whose possible measurement
outcomes are only 0 and 1. The null and the unit observables may be
considered as extreme forms of projectors and are, obviously, the
unique constant projectors. For a generic projector $I$, it always
results:

\begin{equation}
0\leq I\leq1\text{ .}\label{eq:Lim_Indic}
\end{equation}

\noindent It will now prove the following important theorem: \emph{necessary
and sufficient condition in order to an observable $I$ be a projector
is that it is idempotent} \textbf{($I^{2}=I$)}. In fact, if $I$
is a projector, its possible measurement outcomes are $0$ and $1$.
If the outcome of a measurement of $I$ is $0$, also the outcome
for $I^{2}$ will be 0; whereas if the outcome for $I$ is 1, the
same will be for $I^{2}$. So, for the equivalence criterion, it results:

\begin{equation}
I^{2}=I\text{ .}\label{eq:idempotenza}
\end{equation}
Conversely, if the\emph{ idempotency relation} \eqref{eq:idempotenza}
holds, one has:

\begin{equation}
I\left(1-I\right)=0\text{ .}\label{eq:idempot_2}
\end{equation}
In a measurement of the observable in the left-hand side of the \eqref{eq:idempot_2},
due to the zero-product property, that holds for measures, the only
possible outcomes for $I$ are $0$ or $1$. The observable $I$ is
therefore a projector.

We can now demonstrate that if $I$ is a projector, $1-I$ is also
a projector. In fact, it results:

\[
\left(1-I\right){}^{2}=1-2I+I^{2}=1-2I+I=1-I\text{ .}
\]
The projector $1-I$ is the \emph{complementary} projector to $I$.
It is worth noting that if $I$ is not constant, its complementary
too isn't, besides, by virtue of \eqref{eq:idempot_2}, the product
of $I$ and its complementary is null even if no one of the factors
is. This demonstrates that, in the ring of observables, in general
the zero-product property is not valid.

Let's now demonstrate that the product of two compatible projectors,
$I_{1}$ and $I_{2}$, is also a projector. In fact it results:

\begin{equation}
\left(I_{1}I_{2}\right){}^{2}=\left(I_{1}I_{2}\right)\left(I_{2}I_{1}\right)=I_{1}^{2}I_{2}^{2}=I_{1}I_{2}\text{ .}\label{eq:Prod_Ind_Comp}
\end{equation}
Note that the result is true if and only if the commutative and the
associative properties of the product both hold. It is immediate to
verify that \emph{the product of two compatible projectors is always
less or equal to both of them}. Besides, the following property holds:

\begin{equation}
I_{1}\leq I_{2}\;\Rightarrow\;I_{1}I_{2}=I_{1}\text{ .}\label{eq:Prod_Incaps}
\end{equation}
Finally, we will prove that if $I_{1}\leq I_{2}$, the difference:
$I_{2}-I_{1}$ is also a projector. In fact, making use of \eqref{eq:Prod_Incaps},
one has:

\[
\left(I_{2}-I_{1}\right){}^{2}=I_{2}^{2}-2I_{1}I_{2}+I_{1}^{2}=I_{2}-2I_{1}+I_{1}=I_{2}-I_{1}\text{ .}
\]

Two projectors, $I_{1}$ and $I_{2}$, will be called \emph{orthogonal}
if their product is zero:

\begin{equation}
I_{1}I_{2}=0\text{ .}\label{eq:Ind_Alternativi}
\end{equation}
Therefore a projector and its complementary are always orthogonal.
We can demonstrate that\emph{ the sum of two orthogonal projectors
is still a projector}. In fact, it results:

\[
\left(I_{1}+I_{2}\right){}^{2}=I_{1}^{2}+I_{2}^{2}+2I_{1}I_{2}=I_{1}+I_{2}\text{ .}
\]
It is easily verified that both $I_{1}$ and $I_{2}$ are less or
equal to their sum.

A \emph{primitive projector} is defined as a non-null projector that
cannot be expressed as sum of nonzero orthogonal projectors. Primitive
projectors correspond in the Dirac formulation to \emph{pure state
projection operator}, i.e. operators of the form $\ket{\psi}\bra{\psi}$,
where $\ket{\psi}$ is vector state.

It will be proved that a \emph{primitive projector can be major or
equal only of the zero projector}. If, in fact, a primitive projector
$I_{2}$ were major or equal of a nonzero projector, $I_{1}$, for
what above demonstrated, the observable $I_{3}=I_{2}-I_{1}$ would
be a nonzero projector and, besides, by virtue of \eqref{eq:Prod_Incaps},
it would be:

\[
I_{1}I_{3}=I_{1}\left(I_{2}-I_{1}\right)=I_{1}I_{2}-I_{1}^{2}=I_{1}-I_{1}=0\text{ .}
\]
Therefore $I_{2}$ would be equal to the sum of two nonzero orthogonal
projectors, that is a contradiction.

An important consequence of this property of the primitive projectors
is that \emph{the product of a primitive projector for another compatible
projector is or $0$ or equal to the primitive projector itself} (otherwise
one would fall in the previous contradiction).

It is, now, possible to define a \emph{projector basis} $\left\{ I_{j}\right\} $,
as a set of non-null, pairwise orthogonal, primitive projectors that
satisfy the \emph{closure relation}:

\begin{equation}
\sum_{j}I_{j}=1\text{ .}\label{eq:Rel_Chiusura}
\end{equation}

The projectors that form such a basis are \emph{linearly independent}.
A linear combination of them, in fact, using coefficients $\alpha_{j}$,
gives as result an observable whose possible outcomes of measurement
are the coefficients themselves. This, therefore, will be equal to
the null observable if and only if all the coefficients $\alpha_{j}$
are equal to zero.

\subsection{Projectors and events\label{subsec:Projectors-and-Events}}

In order to start to clarify the physical meaning of the projectors,
it is helpful to introduce the concept of \emph{event associated projector}.
We will say that $I_{\mathcal{E}}$ is the projector associated to
the event $\mathcal{E}$, if the outcome of measurement of $I_{\mathcal{E}}$
is equal to $1$ if and only if the event $\mathcal{E}$ occurs. With
this definition, one can state that the zero observable is the projector
associated to the \emph{impossible event}, whereas the unit observable
is the projector associated to the \emph{certain event}.

The complementary projector is, moreover, associated with the event
complementary of that associated to the starting projector and orthogonal
projectors are associated to mutually exclusive events, i.e. such
that the occurrence of one precludes the occurrence of the others.
An important consequence of this last association is that, therefore,
\emph{pairwise orthogonal projectors are compatibles among each other},
since the verify on the events implies the simultaneous measuring
of all of the projectors.

The sum of orthogonal projectors is then associated to the \emph{union}
of the events associated to each of them, the compound event which
occurs if one of these events occurs.

The product of two compatible projectors gives as result the projector
associated to the \emph{intersection} of the events associated to
the factors, namely the event that occurs when both events occur.

The primitive projectors, finally, are associated to \emph{elementary
events}, not reducible to the union of simpler events.

\subsection{Observable decomposition\label{subsec:Scomposizione_Osservabile}}

Now let's consider an observable $O$, having the spectrum$\left\{ o_{j}\right\} $.
If we denote with $O=o_{j}$ the event that occurs when by measuring
$O$ one obtains $o_{j}$ as outcome, it can be proved that it results:

\begin{equation}
O=\sum_{j}o_{j}\:I_{O=o_{j}}\text{ .}\label{eq:I_Scomp_Oss}
\end{equation}
The outcomes of measurement of the observables in both sides of \eqref{eq:I_Scomp_Oss},
indeed, coincide in all measurements. In fact, if, by measuring $O$,
one obtains a certain outcome $o_{k},$ the measurement of the projectors
in the right-hand side of \eqref{eq:I_Scomp_Oss} will give 0 as outcome
except for the projector $I_{O=o_{k}}$, for which one will get $1$,
and therefore the outcome of measurement of the sum will be equal
to: $o_{k}\:1+0=o_{k}$.

Note that the projectors $I_{O=o_{j}}$ are pairwise orthogonal and
satisfy the closure relation \eqref{eq:Rel_Chiusura}, since the union
of the events associated gives as result the certain event (the measurement
of $O$ must have as outcome one of $o_{j}$). These, therefore, form
a projector basis uniquely determined, apart from the order, by the
observable $O$. We will call it projector basis \emph{associated
}to the observable $O$. Notice, however, that, generally speaking,
the projectors of this basis aren't primitive.

Consider, now, two compatible observables $O_{1}$ and $O_{2}$. According
to what above proved, they may be written in the form:

\begin{eqnarray*}
O_{1} & = & \sum_{j}o_{1,j}\:I_{O_{1}=o_{1,j}}\text{ ,}\\
O_{2} & = & \sum_{k}o_{2,k}\:I_{O_{2}=o_{2,k}}\text{ .}
\end{eqnarray*}
One should observe that the compatibility of $O_{1}$ and $O_{2}$
implies that also the projectors $I_{O_{1}=o_{1,j}}$ and $I_{O_{2}=o_{2,k}}$,
that appear in the two above decompositions, must be compatible among
themselves, since their measurability depends on that of $O_{1}$
and $O_{2}$, respectively.

By applying the closure relation \eqref{eq:Rel_Chiusura} to the projector
bases associated to the two observables, one obtains:

\begin{eqnarray*}
O_{1} & =O_{1}1= & \sum_{j}o_{1,j}\:I_{O_{1}=o_{1,j}}\sum_{k}I_{O_{2}=o_{2,k}}=\sum_{j,k}o_{1,j}\:I_{O_{1}=o_{1,j}}\,I_{O_{2}=o_{2,k}}\text{ ,}\\
O_{2} & =O_{2}1= & \sum_{k}o_{2,k}\:I_{O_{2}=o_{2,k}}\sum_{j}I_{O_{1}=o_{1,j}}=\sum_{j,k}o_{2,k}\:I_{O_{1}=o_{1,j}}\,I_{O_{2}=o_{2,k}}\text{ .}
\end{eqnarray*}
It is, therefore, possible to express both the observables as linear
combinations of the same set of projectors:

\[
I_{O_{1}=o_{1,j}}\,I_{O_{2}=o_{2,k}}=I_{O_{1}=o_{1,j}\wedge O_{2}=o_{2,k}}\text{ .}
\]
The projectors of this set are clearly pairwise orthogonal and satisfy
the closure relation:

\[
\sum_{j,k}I_{O_{1}=o_{1,j}\wedge O_{2}=o_{2,k}}=\sum_{j,k}I_{O_{1}=o_{1,j}}\,I_{O_{2}=o_{2,k}}=\sum_{j}I_{O_{1}=o_{1,j}}\sum_{k}I_{O_{2}=o_{2,k}}=1\text{ .}
\]
They thus form a new projector basis.

In general, if $\left\{ I_{j}\right\} $ is a projector basis and
$O$ is an observable that can be put in the form of a linear combination
of them of the type:

\begin{equation}
O=\sum_{j}o_{j}\,I_{j}\label{eq:Scomp_Oss}
\end{equation}
where the coefficients $o_{j}$ aren't necessarily all distinct, we
will call the \eqref{eq:Scomp_Oss} \emph{decomposition of the observable
$O$ according to the projector basis }\textbf{\textit{\emph{$\left\{ I_{j}\right\} $}}}.
The coefficients $o_{j}$, will be called, besides, \emph{spectral
coefficients of the observable $O$ with respect to the projector
basis }\textbf{\textit{\emph{$\left\{ I_{j}\right\} $}}}. Clearly,
the spectral coefficients of an observable are the same numbers of
its possible measurement outcomes.

By multiplying both sides of the \eqref{eq:Scomp_Oss} for a generic
projector $I_{j_{0}}$ of the basis, one obtains the following important
relation:

\begin{equation}
OI_{j_{0}}=o_{j_{0}}I_{j_{0}}\label{eq:Eq_autoval}
\end{equation}
which characterizes the value of the spectral coefficient $o_{j_{0}}$
of a given observable $O$, relative to the projector $I_{j_{0}}$.

By adding further compatible observables, therefore, one obtains sets
of pairwise orthogonal projectors, all satisfying the closure relation
and always minor or equal to the starting ones. The procedure comes
to an end when all the projectors obtained are primitive: at this
point the addiction of a further observable can no more alter the
set of projectors.

The set, that will be assumed as \emph{finite}, of observables $\left\{ O_{r}\right\} $,
independent of each other, that are needed to individuate, by means
of the described procedure, a basis of pairwise orthogonal primitive
projectors, will be called a \emph{complete set of compatible observables}\footnote{The analogies with the complete sets of \emph{commuting} observables,
introduced by Dirac in his book \citep{Dirac1930}, page 57, are clear.
One should, however, note the difference in the point of view: here
there is no reference to eigenstates.}. The above outlined procedure proves that \emph{every observable
is an element of at least one complete set of compatible observables}.
If, in fact, the projectors of the basis associated to an observable
$O$ were not primitive, they, by definition, would be expressible
as a sum of nonzero pairwise orthogonal, primitive projectors, each
of which would be compatible with $O$ (in fact $O$ results to be
a linear combination of them and other projectors that are mutually
exclusive with them). The set of compatible observables formed by
$O$ and by such projectors would be, therefore, complete.

Note, besides, that if $I_{j}$ and $I_{k}$ ($j\neq k$) are two
distinct primitive projectors of the basis, according to the basis
of pairwise orthogonal, primitive projectors construction procedure,
the set of coefficients $\left\{ o_{r,j}\right\} $ associated to
the first projector and that of those, $\left\{ o_{r,k}\right\} $,
associated to the second must differ for at least one element.

We can now define a \emph{function of an observable} $O$ as an observable
whose measurement have as outcomes the images, under a certain function
$f$, of the outcomes of measurement of the observable $O$. It can,
therefore, be expressed, in terms of the projector basis associated
to \emph{$O$}, in the form:

\begin{equation}
f\left(O\right)\ideq\sum_{j}f\left(o_{j}\right)I_{O=o_{j}}\text{ .}\label{eq:Def_Funzione_Oss}
\end{equation}
More generally, if an observable $O$ is decomposed according to a
projector basis $\left\{ I_{j}\right\} $, in the form of \eqref{eq:Scomp_Oss},
it immediately follows by definition, making reference to \eqref{eq:Def_Funzione_Oss},
that it results:

\begin{equation}
f\left(O\right)=\sum_{j}f\left(o_{j}\right)I_{j}\label{eq:Def_Funz_Oss_Gen}
\end{equation}
for every function of the observable $O$. Clearly \emph{all functions
of an observable are compatible among each other}.

It is worth noting that, by defining the \emph{Kronecker $\delta$
function} as usual:

\begin{equation} \delta(x)\ideq\cases {0 & \text{for} $x\neq0$\\ 1 & \text{for} $x=0$ }\label{eq:Def_Funz_Kronecker} \end{equation}chosen
an index $j$, one has:
\begin{equation}
I_{O=o_{j}}=\delta\left(O-o_{j}\right)\text{ .}\label{eq:Indicat_Funz}
\end{equation}
So\emph{ the projectors of a basis associated to a certain observable
may be considered as functions of the observable itself}. Projectors
of bases associated to compatible observables, therefore, are compatible
among each other. This proves what stated before.

The definition of a function of an observable, given above, can easily
be extended also to the case of functions of several observables compatible
among each other. If one denotes with $\boldsymbol{O}$ the $n$-tuple
of the compatible observables:

\[
\boldsymbol{O}\ideq\left(O_{1},\,\ldots,\,O_{r},\,\ldots\right)
\]
and with$\boldsymbol{o}_{j}$ the $n$-tuple of the corresponding
coefficients associated to the primitive projector $I_{j}$ of the
basis:
\[
\boldsymbol{o}_{j}\ideq\left(o_{1j},\,\ldots,\,o_{rj},\,\ldots\right)
\]
we can define a function $f$ of the observables $\boldsymbol{O}$
as:

\begin{equation}
f\left(\boldsymbol{O}\right)\ideq\sum_{j}f\left(\boldsymbol{o}_{j}\right)I_{j}\text{ .}\label{eq:Funzione_Osservabili}
\end{equation}

If an observable $A$ is compatible with a complete set of compatible
observables, that individuates a basis $\left\{ I_{j}\right\} $ of
pairwise orthogonal, primitive projectors, for what outlined before,
it will be decomposable, according to the projector basis $\left\{ I_{j}\right\} $,
in the form \eqref{eq:Scomp_Oss}, with spectral coefficients $a_{j}$.
We will now prove that \emph{the observable $A$ is a function of
the observables }\textbf{\emph{$\boldsymbol{O}$ }}\emph{of the complete
set}. In fact, after observing that, by changing the index $j$, the
$n$-tuple $\boldsymbol{o}_{j}$ are distinct among each other, it
suffices to put: $f\left(\boldsymbol{o}_{j}\right)\ideq a_{j}$.

We will call a \emph{complete space of compatible observables}, a
space of compatible observables containing a complete set of compatible
observables. For them it holds the following fundamental \emph{theorem
of uniqueness of the basis of pairwise orthogonal, primitive projectors}:
\emph{every complete space of compatible observables admits an unique
basis of pairwise orthogonal, primitive projectors}. Let's suppose,
in fact, for absurd, that a complete space of compatible observables
admits two distinct bases of pairwise orthogonal, primitive projectors:
$\left\{ I_{j}\right\} $ and $\left\{ J_{k}\right\} $. For a generic
projector $I_{j_{0}}$ of the first basis, one would have:

\[
I_{j_{0}}=I_{j_{0}}1=I_{j_{0}}\sum_{k}J_{k}=\sum_{k}I_{j_{0}}J_{k}\text{ .}
\]
Since $I_{j_{0}}$ is primitive, the products in the last summation
should cancel out all but one, corresponding to a certain index $k_{0}$,
and so it should result $I_{j_{0}}=I_{j_{0}}J_{k_{0}}$. But being
$J_{k_{0}}$ also primitive, the product $I_{j_{0}}J_{k_{0}}$, not
being equal to $0$, should be equal to $J_{k_{0}}$, and so, by transitivity,
it should result: $I_{j_{0}}=J_{k_{0}}$. Every projector of the first
basis, therefore, will be equal to one of the second and, by symmetry,
every projector of the second basis too will be equal to one of the
first. The two bases will so be identical.

For what above proved, every observable $O$, belonging to a complete
space of compatible observables with the basis $\left\{ I_{j}\right\} $
of pairwise orthogonal, primitive projectors, will be decomposable,
according to this basis, in the form \eqref{eq:Scomp_Oss}. This decomposition,
besides, is \emph{unique}, in force of the theorem of uniqueness of
the basis of pairwise orthogonal, primitive projectors.

We will define \emph{multiplicity} of a given outcome of measurement
$o$ of an observable $O$, the number of times that this coefficient
appears in the decomposition according to the basis of primitive projectors.

It is worth, now, however, to point out an important clarification.
The circumstance that a given set of independent compatible observables,
relative to a certain physical system, is \emph{complete}, is, actually,
a \emph{characterization} of the representation that one assume to
adopt for the physical system itself. It is therefore more correct
to state that the assumption that a given set of compatible observable
is complete characterize a \emph{model} of the physical system in
study, model that, if necessary, may be also subsequently amended
by adding new independent observable to the starting ``complete''
set.

\section{Incompatible Observables}

\subsection{The issue of the incompatible observables}

The existence of incompatible observables complicates the conceptual
framework introduced so far, but, at the same time, enriches it also
greatly.

First of all, it is to be noted that the compatibility between observables
is not an equivalence relation, since, in general, it is not valid
the transitive property. However, different complete spaces of compatible
observables have always the constants as common elements. The set
of observables belonging to both spaces forms the \emph{intersection
space}. The intersection space contains the constants and it is therefore
definitely not empty. However, it does not admit a basis of pairwise
orthogonal, primitive projectors, except in the trivial case in which
the two complete spaces coincide.

The introduction of operations between incompatible observables is
far from being trivial. The underlying issue is that you cannot even
give an adequate operational definition of sum or product, in the
sense that it is not possible to express the results of these operations
solely in terms of the spectra of the observables themselves. The
reason for this difficulty lies in the concept of non-compatibility
of observables: not being possible a simultaneously measure the two
observables on the same ``copy'' (ensemble member) of the system
in the given state, it is not permissible to combine the results of
two successive measurements for obtain the measurement of a new quantity,
as the second measurement is influenced by the information arising
from the first (\emph{conditioned events}).

The operations involving incompatible observables shall, therefore,
be treated by following a different approach. The aim is to define
them, in a such a way that, in the case of compatible observables,
the sum and the product are reduced to what previously defined. In
general, however, the result will be only a \emph{formal expression}
corresponding to a non-observable entity, that we will call \emph{pseudo-observable}.

In defining the operations, we will try to preserve, as much as possible,
their fundamental properties. In particular, the operations will be
defined in such a way that the set of the pseudo-observables retains,
with respect to these operations, the algebraic structure of \emph{vector
space} and \emph{unitary ring}. It will be assumed, therefore, that
for the addition the commutative property, the associative property,
the existence of the additive identity (obviously the constant $0$)
and existence of the additive inverse all hold. For the multiplication,
we will assume that they hold the associative property, \emph{but
not the commutative one}, and both the left and right distributivity
over addition. The constant $1$, besides, will be the multiplicative
identity.

To give up the commutative property is necessary in order to preserve
the possibility that two observables may be incompatible. Let, in
fact, be $A$ and $B$ two incompatible observables. Let, besides,
be $\mathbb{A}$ a complete space, having the basis $\left\{ I_{\mathbb{A},j}\right\} $
of pairwise orthogonal, primitive projectors, containing the observable
$A$ and be $\mathbb{B}$ a complete space, with the basis $\left\{ I_{\mathbb{B},k}\right\} $
of pairwise orthogonal, primitive projectors, containing the observable
$B$. If the product of two observables were always commutative and
gave as result another observable, the product of primitive projectors
$I_{\mathbb{A},j}\,I_{\mathbb{B},k}$, according to \eqref{eq:Prod_Ind_Comp},
would be a projector too. Following the same lines of reasoning used
to prove the theorem of uniqueness of the basis of pairwise orthogonal,
primitive projectors, one would conclude that the bases $\left\{ I_{\mathbb{A},j}\right\} $
and $\left\{ I_{\mathbb{B},k}\right\} $ should coincide and therefore
all their linear combination, included the observables $A$ and $B$,
should be compatible, which is a contradiction.

\subsection{Hermitian transposition of pseudo-observables\label{subsec:Trasposizione}}

But what physical meaning has to be given to an operation involving
incompatible observables? Since two incompatible observables cannot
be measured simultaneously, the \emph{observation order} must be relevant.
The term ``observation order'' here refers to the sequence in which
the observer plans to perform measurements of the various observables
under investigation, which is obviously relevant in the case of incompatible
observables. It is worth to point out that the observation order is
independent of the eventual temporal ordering of the observables,
since it is experimentally possible to measure \emph{before} an observable
relative to a time posterior with respect to the others, by recording,
for instance exploiting the \emph{entanglement} mechanism, the information
in different instants of time for a subsequent measurement in a whatever
order.

We will then adopt the convention that \emph{in an operation the rightmost
observable has to be measured before the leftmost}.

It is besides clear that an observable, defined through operations
on other observables, does not change if one inverts the observation
order.

With these ideas in mind, we can now introduce the \emph{Hermitian
transposition involution} ($\dagger$)\footnote{The reference to the Hermitian conjugate of the operators associated
to the observables in the Dirac-von Neumann formulation of quantum
mechanics is obvious.}, which switches from an expression to that obtained by reversing
the observation order of the observables involved in it. The Hermitian
transposition involution, acting on the \emph{space of the pseudo-observables}
$\mathbb{P}$, is defined by the following properties:
\begin{enumerate}
\item For each pseudo-observable $P$ one has:$\left(P^{\dagger}\right)^{\dagger}=P$.
\item A pseudo-observable $P$ is an observable iff $P^{\dagger}=P$ (\emph{maximum
physical significance assumption}).
\item For each pair of pseudo-observables $P$ and $Q$ one has: $\left(P+Q\right)^{\dagger}=P^{\dagger}+Q^{\dagger}$.
\item For each pair of pseudo-observables $P$ and $Q$ one has: $\left(PQ\right)^{\dagger}=Q^{\dagger}P^{\dagger}$.
\item For each pseudo-observable $P$ one has: $PP^{\dagger}\geq0$.
\item For a pseudo-observable $P$ one has $PP^{\dagger}=0$ iff $P=0$.
\end{enumerate}
The first property follows from the fact that reversing twice the
observation order one comes back to the starting situation.

The second property gives the maximum possible physical meaning to
the transposition, by connecting it directly to the observability
of a given expression.

The third and fourth properties derive directly from the definitions
and from the assumption that the sum of observables is commutative.

The fifth property extends to pseudo-observables the \emph{condition
of reality of an observable}, according to which the square of an
observable is always positive or zero. The expression to the first
member is constructed in such a manner to ensure that it is an observable.

The sixth property is the extension to the case of pseudo-observables
of a similar property on the square of the observables, easily proved
by considering the possible outcomes of measurement.

The Hermitian transposition operator allows us to generalize to the
case of pseudo-observables the idempotency relation \eqref{eq:idempotenza},
characteristic of the projectors. In fact, it is easy to prove that
necessary and sufficient condition in order a pseudo-observable $I$
is a projector is that it results:
\begin{equation}
II^{\dagger}=I\text{ .}\label{eq:Cond_Idempotenza_Gen}
\end{equation}
Indeed, if $I$ is a projector, due to the second property of the
Hermitian transposition, the \eqref{eq:Cond_Idempotenza_Gen} is equivalent
to \eqref{eq:idempotenza}. If, instead, $I$ satisfies the \eqref{eq:Cond_Idempotenza_Gen},
$I$ is an observable, being the left hand side invariant for Hermitian
transposition, and so it satisfies the idempotency relation \eqref{eq:idempotenza},
to which \eqref{eq:Cond_Idempotenza_Gen} reduces itself in the case
of observables. $I$ is therefore a projector.

\subsection{Complex form of the pseudo-observables\label{subsec:FormaComplessaPO}}

Let's consider now a generic pseudo-observable $P$, generated by
addition and multiplication of compatible and incompatible observables.
Note that it results:

\begin{equation}
P=P_{\mathrm{S}}+P_{\mathrm{A}}\qquad\mbox{with: }P_{\mathrm{S}}=\frac{P+P^{\dagger}}{2}\mbox{ and }P_{\mathrm{A}}=\frac{P-P^{\dagger}}{2}\label{eq:Forma_SA_PO}
\end{equation}
where, since $P_{\mathrm{S}}^{\dagger}=P_{\mathrm{S}}$, $P_{\mathrm{S}}$
is an observable, while $P_{\mathrm{A}}^{\dagger}=-P_{\mathrm{A}}$.
$P_{\mathrm{S}}$ is the \emph{Hermitian part} of the pseudo-observable
$P$, while $P_{\mathrm{A}}$is its \emph{skew-Hermitian part}.

In order to give to the pseudo-observables a much more meaningful
form than \eqref{eq:Forma_SA_PO}, we now introduce a nonzero pseudo-observable
$i$, whose product with any other pseudo-observable is commutative
and such that it results:

\begin{equation}
i^{\dagger}=-i\label{eq:Trasp_i}
\end{equation}
or, in other words, whose Hermitian part is equal to the zero observable.

To fully define the pseudo-observable $i$, we first observe that,
for the fifth and the sixth properties of the Hermitian transposition
operator, it results:

\[
ii^{\dagger}=-i^{2}>0\;\Rightarrow\;i^{2}<0
\]
so that $i^{2}$ turns out to be a negative observable (for ``negative
observable'' we mean an observable whose possible outcomes of measurement
are all negative). One is, therefore, allowed to put:
\begin{equation}
i^{2}\ideq-1\text{ .}\label{eq:i2}
\end{equation}

The pseudo-observable $i$, then, behaves like the imaginary unit
of the complex numbers, with which, in what it follows, it will be
identified.

It is now worth observing that, for the skew-Hermitian $P_{\mathrm{A}}$
of a pseudo-observable $P$, it results:

\begin{equation}
\left(-iP_{\mathrm{A}}\right)^{\dagger}=-i^{\dagger}P_{\mathrm{A}}^{\dagger}=i\left(-P_{\mathrm{A}}\right)=-iP_{\mathrm{A}}\label{eq:Propr_Parte_Imm}
\end{equation}
which proves that the product of the skew-Hermitian part of a pseudo-observable
$P$ and the opposite of $i$ is an observable, that will be called
the \emph{imaginary part}, $P_{\mathrm{I}}$, of $P$\emph{:}

\begin{equation}
\imp P\ideq-iP_{\mathrm{A}}\text{ .}\label{eq:Def_Parte_Imm}
\end{equation}
Making use of \eqref{eq:Def_Parte_Imm}, \eqref{eq:Forma_SA_PO} and
\eqref{eq:i2}, we can finally give the following \emph{complex representation}
of a pseudo-observable:

\begin{equation}
P=P_{\mathrm{S}}+P_{\mathrm{A}}=\rp P+i\imp P\label{eq:Forma_Compl_PO}
\end{equation}
where we have put $\rp P\ideq P_{\mathrm{S}}$, the \emph{real part}
of $P$.

The introduction of the pseudo-observable $i$, according to \eqref{eq:Forma_Compl_PO},
allows to associate every pseudo-observable to a pair of observables,
its real and imaginary part. In this manner the unique non-physical
entity needed to express a pseudo-observable is just the imaginary
unit\emph{ i}. It is also necessary, since if the imaginary part of
each pseudo-observable were the zero observable, then every of them,
reducing itself to its Hermitian part, should be an observable. In
particular, in such hypothesis, the product of each pair of observables
would be an observable. But this would imply, by virtue of the fourth
and the second properties of the Hermitian transposition operator,
that every pair of observables commutes. But in this case, as it will
be demonstrated in section ``Transformations'' of the third paper,
no transformation would be allowed and therefore no evolution: an
Universe of only compatible observables would be static, definitely
dead! Therefore \emph{an evolving, living Universe needs to be described
by complex representable pseudo-observables}.

According to \eqref{eq:Forma_Compl_PO} and \eqref{eq:Trasp_i}, one
has:

\begin{equation}
P^{\dagger}=\left(\rp P+i\imp P\right)^{\dagger}=\rp P-i\imp P\label{eq:Compl_Con_P}
\end{equation}
that proves the analogous of a well known property of the \emph{Hermitian
conjugate }of an operator in the Dirac formulation of quantum mechanics.

In what follows, we will call \emph{complex constants} the pseudo-observables
of the form:

\begin{equation}
\gamma=\alpha+i\beta\label{eq:Pseudocostanti}
\end{equation}
where $\alpha$ and $\beta$ are two constant observables. They will
be identified with their complex value, i.e. with the complex number
having the value of their real part as real part and the value of
their imaginary part as imaginary part. In such a context, we will
refer to constant observables with the term of \emph{real constants}.

By this definition, for each complex constant $\gamma$, it results:

\begin{equation}
\gamma^{\dagger}=\gamma^{*}\label{eq:Complesso_Coniugato_PC}
\end{equation}
where the star ($*$) indicates the \emph{complex conjugate}.

The introduction of the pseudo-observable $i$ allows us to define
the set $\mathbb{P}$ of the pseudo-observables as that formed by
every expression of the form:
\[
P\ideq A+iB
\]
where $A$ and $B$ are two generic observables (compatible or incompatible).
For what above discussed, this set, with the operations of addition
and multiplication, has the algebraic structure of a ring.

\subsection{Commutator and compatibility\label{subsec:Commutatori}}

As already seen, the non-commutativity of the product of two observables
is tied in a deep way to their incompatibility.

In order to make this point more clear, it is useful to introduce
the \emph{commutator} of two observables $A$ and $B$, defined, as
well known, as follows:

\begin{equation}
\left[A,B\right]\ideq AB-BA\text{ .}\label{eq:Def_Commutatore}
\end{equation}
This definition can be extended, in an obvious manner, to the more
general case of two pseudo-observables.

If $A$ and $B$ are two generic pseudo-observables, it is immediate
to prove the following properties of the commutator: 
\begin{enumerate}
\item The commutator is \emph{anti-commutative}: $\left[A,B\right]=-\left[B,A\right]$.
\item The commutator is \emph{nilpotent}: $\left[A,A\right]=0$.
\item The commutator is a \emph{bilinear form}:\\
$\left[c_{1}\,A+c_{2}\,B,C\right]=c_{1}\left[A,C\right]+c_{2}\left[B,C\right]$
\\
$\left[A,c_{1}\,B+c_{2}\,C\right]=c_{1}\left[A,B\right]+c_{2}\left[A,C\right]$\\
where $c_{1}$ and $c_{2}$ are two generic constants (real or complex).
\item The commutator satisfies the following forms of the \emph{Leibniz
rule}:\\
$\left[A,BC\right]=\left[A,B\right]C+B\left[A,C\right]$\\
$\left[AB,C\right]=A\left[B,C\right]+\left[A,C\right]B\text{ .}$
\item The commutator satisfies the following \emph{Jacobi identity}:\\
 $\left[A,\left[B,C\right]\right]+\left[B,\left[C,A\right]\right]+\left[C,\left[A,B\right]\right]=0\text{ .}$
\end{enumerate}
In analogy with the Dirac formulation, we will say that two observables,
or pseudo-observables, \emph{commute} if their commutator is equal
to $0$.

Consider two generic observables, $A$ and $B$, which commute with
each other: $\left[A,B\right]=0$. We will prove that it results $\left[A^{n},B\right]=0$
for every natural exponent $n$. This is trivially true for $n=1.$
Assumed true for the exponent $n-1$, moreover one has:
\[
\left[A^{n},B\right]=\left[AA^{n-1},B\right]=A\left[A^{n-1},B\right]+\left[A,B\right]A^{n-1}=0\text{ .}
\]
The property is so demonstrated by induction. Applying the same reasoning
to the powers of $B$, more generally, one obtains:

\[
\left[A^{n},B^{m}\right]=0\qquad\forall n,m\in\mathbb{N\text{ .}}
\]
Besides, due to the bilinearity of the commutator, each polynomial
expression in $A$ commutes with any polynomial expression in $B$,
i.e. if $p$ and $q$ are two whatever polynomials, one has:

\begin{equation}
\left[A,B\right]=0\;\Rightarrow\;\left[p\left(A\right),q\left(B\right)\right]=0\text{ .}\label{eq:Commutazione_Polinomi}
\end{equation}

Having premised that, we now can demonstrate that \emph{necessary
and sufficient condition in order that two observables are compatible
is that they commute}.

We begin to prove that two compatible observables commute. Let $A$
and $B$ two compatible observables and so belonging to the same complete
space of compatible observables, having the basis $\left\{ I_{j}\right\} $
of primitive projectors. By decomposing the observables $A$ and $B$
according to this basis, one has:

\[
A=\sum_{j}a_{j}\,I_{j}\qquad\mbox{and}\qquad B=\sum_{k}b_{k}\,I_{k}
\]
and therefore for \eqref{eq:idempotenza} and \eqref{eq:Ind_Alternativi}:
\[
AB=\sum_{j,k}a_{j}b_{k}\,I_{j}I_{k}=\sum_{j}a_{j}b_{j}\,I_{j}=\sum_{j,k}b_{k}a_{j}\,I_{k}I_{j}=BA
\]
that implies: $\left[A,B\right]=0$.

The proof of the sufficiency of the condition is more complex. Suppose,
therefore, that $A$ and $B$ are two observables that commute, and
therefore such that it results:

\begin{equation}
\left[A,B\right]=0\text{ .}\label{eq:Commutazione}
\end{equation}
By decomposing these observables according to the bases associated
with them, for the \eqref{eq:I_Scomp_Oss}, one has:

\noindent
\begin{equation}
A=\sum_{j}a_{j}\,I_{A=a_{j}}\qquad\mbox{and}\qquad B=\sum_{k}b_{k}\,I_{B=b_{k}}\label{eq:Scomp_AeB}
\end{equation}
where both the coefficients $a_{j}$ and $b_{k}$ are distinct among
themselves.

Since the observables $A$ and $B$ commute, by virtue of the property
\eqref{eq:Commutazione_Polinomi}, the \eqref{eq:Scomp_AeB} and \eqref{eq:Funzione_Osservabili},
considered two whatever polynomial expressions $p\left(A\right)$
and $q\left(B\right)$, respectively in $A$ and $B$, it results:

\noindent
\begin{equation}
\left[p\left(A\right),q\left(B\right)\right]=\sum_{j,k}p\left(a_{j}\right)q\left(b_{k}\right)\left[I_{A=a_{j}},I_{B=b_{k}}\right]=0\text{ .}\label{eq:Rel_Comm_Polinomi}
\end{equation}

\noindent Chosen arbitrarily two indexes, $j'$ and $k'$, we now
put:

\noindent
\begin{equation}
p\left(A\right)=\prod_{j\neq j'}\left(A-a_{j}\right)\qquad\text{and}\qquad q\left(B\right)=\prod_{k\neq k'}\left(B-b_{k}\right)\text{ .}\label{eq:Polinomi_AB}
\end{equation}

\noindent Note that, by construction, one has $p\left(a_{j}\right)=0$
for each index $j\neq j'$ and $q\left(b_{k}\right)=0$ for each index
$k\neq k'$. It also results: 

\noindent
\[
p\left(a_{j'}\right)=\prod_{j\neq j'}\left(a_{j'}-a_{j}\right)\neq0
\]
since the coefficients $a_{j}$ are all each other distinct. For the
same reason, it also holds: $q\left(b_{k'}\right)\neq0$. By applying
these relations to \eqref{eq:Rel_Comm_Polinomi} and exploiting the
arbitrariness in the choice of the indexes $j'$ and $k'$, one can
therefore state that, for each pair of indexes $j$ e $k$, it has
to result:

\noindent
\begin{equation}
\left[I_{A=a_{j}},I_{B=b_{k}}\right]=0\text{ .}\label{eq:Commut_Indic}
\end{equation}

But if two projectors commute, their product is an observable and
is still a projector. It is then easy to verify that the set of the
products of projectors thus obtained satisfies the relation of closure
\eqref{eq:Rel_Chiusura} and consists of pairwise orthogonal projectors,
for which it holds \eqref{eq:Ind_Alternativi}. According to the link
between projectors and events, therefore, these projectors are each
other compatible, since they are associated with events mutually exclusive
and therefore necessarily observable simultaneously, and form a basis.

According to \eqref{eq:Scomp_AeB} and to the closure relation \eqref{eq:Rel_Chiusura},
it then follows:

\begin{eqnarray*}
A & = & A\,1=\sum_{j}a_{j}\,I_{A=a_{j}}\sum_{k}I_{B=b_{k}}=\sum_{j,k}a_{j}\,I_{A=a_{j}}I_{B=b_{k}}\text{ ,}\\
B & = & 1\,B=\sum_{j}I_{A=a_{j}}\sum_{k}b_{k}\,I_{B=b_{k}}=\sum_{j,k}b_{k}\,I_{A=a_{j}}I_{B=b_{k}}\text{ .}
\end{eqnarray*}
The observables $A$ and $B$, therefore, being expressible as linear
combinations of the projectors of the same base, are each other compatible.

An important consequence of the theorem just proved is that \emph{if
two observables commute, also every function of one commutes with
whatever function of the other}. If, in fact, the two observables
commute each other, they are both expressible as a linear combination
of the projectors obtained by multiplying the projectors of the bases
associated with the two observables and discarding the zero products.
These projectors form a basis. Each function of the two observables,
therefore, for \eqref{eq:Def_Funz_Oss_Gen}, will be expressible as
a linear combination of the projectors of the same base, and then
each pair of functions of the two observables will be constituted
by compatible observables which commute among themselves.

Another important consequence is given by the following theorem:\emph{
if an observable commutes with the elements of a basis of pairwise
orthogonal primitive projectors, then the observable is expressible
as a linear combination of the projectors of the basis}. If, in fact,
an observable $A$ commutes with the elements of a basis $\left\{ I_{k}\right\} $
of pairwise orthogonal, primitive projectors, after having written
the observable $A$ in the form of the first of the \eqref{eq:Scomp_AeB},
it should first be noted that also every polynomial expression in
$A$, in particular those of the form of the first of the \eqref{eq:Polinomi_AB},
commutes with each projector of the basis. This implies, as seen before,
that also every projector $I_{A=a_{j}}$, which appears in the decomposition
of $A$, commutes with each projector $I_{k}$ of the basis. The product
$I_{A=a_{j}}I_{k}$ is therefore a projector and, for the properties
of the primitive projectors, it is or zero or equal to $I_{k}$ and
orthogonal to all other products. The thesis then follows by observing
that it results: 

\[
A=A\,1=\sum_{j}a_{j}\,I_{A=a_{j}}\sum_{k}I_{k}=\sum_{k}a_{j_{k}}I_{k}
\]
where $j_{k}$ is the only index for which one has:

\[
I_{A=a_{j_{k}}}I_{k}=I_{k}\text{ .}
\]

It is, finally, worth noting that the imaginary part of the product
of two observables $A$ and $B$:

\begin{equation}
\imp{\left(AB\right)}=\frac{1}{i}\left[A,B\right]\label{eq:Oss_Commut}
\end{equation}
is an observable which, for what above proved, gives a measure of
their grade of incompatibility.

\section{Dyads}

\subsection{Projections and Dyadic Forms\label{subsec:Forme diadiche}}

The set $\mathbb{P}$ of the pseudo-observables constitutes, with
respect to the operations of addition and multiplication by a constant,
a vector space over the field of complex numbers $\mathbb{C}$. In
this space, by virtue of \eqref{eq:Complesso_Coniugato_PC} and of
the properties of the Hermitian transposition operator, for each pair
of constants $\gamma_{1}$ and $\gamma_{2}$ and for each pair of
pseudo-observables $Z_{1}$ and $Z_{2}$, the following relation holds:

\begin{equation}
\left(\gamma_{1}Z_{1}+\gamma_{2}Z_{2}\right)^{\dagger}=\gamma_{1}^{*}Z_{1}^{\dagger}+\gamma_{2}^{*}Z_{2}^{\dagger}\label{eq:Antilinerita}
\end{equation}
which is expressed by saying that the Hermitian transposition operator
is \emph{antilinear}.

Consider a space of compatible observables having the basis$\left\{ I_{j}\right\} $
of pairwise orthogonal primitive projectors. We will show how it is
possible to construct a set of generators for the space $\mathbb{P}$
starting from the projectors $\left\{ I_{j}\right\} $. In order to
do this, one has to observe that, if $P$ is a whatever pseudo-observable,
it results:

\begin{equation}
P=1\,P\,1=\sum_{j,k}I_{j}PI_{k}\text{ .}\label{eq:Generatori_P}
\end{equation}
According to \eqref{eq:Generatori_P}, any pseudo-observable $P$
is then given by a linear combination of the pseudo-observables obtained
by multiplying the pseudo-observable to the right and to the left
for an element of the basis of pairwise orthogonal primitive projectors.

We now consider a product of the form: $I_{j}CI_{j}$, where $C$
is an observable, that will be called, for a reason that will be explained
shortly, \emph{projection} of $C$ according the projector $I_{j}$.
It is immediate, first of all, to prove that this expression is an
observable that commutes with each of the primitive projectors of
the basis $\left\{ I_{k}\right\} $. By the theorem proved at the
end of subsection \ref{subsec:Commutatori}, the projection is expressible
as a linear combination of the projectors of the basis $\left\{ I_{k}\right\} $.
Applying the associative property of the product of observables and
the fact that the projectors of the basis are pairwise orthogonal,
it is concluded finally that the projection of $C$ according the
projector $I_{j}$ is proportional to the only projector $I_{j}$:

\begin{equation}
I_{j}CI_{j}=c_{j}I_{j}\label{eq:Esp_Proiezione_j}
\end{equation}
being $c_{j}$ a suitable real constant, which we will call the \emph{component
of the observable $C$ according the primitive projector $I_{j}$}.
It is possible to give a physical interpretation to this formula,
by noting that the measurement of the projection of $C$ according
the projector $I_{j}$ gives as outcome $c_{j}$ when it occurs the
event associated to the projector $I_{j}$. In such a case the measurement
of this projector gives as outcome 1. According to \eqref{eq:Esp_Proiezione_j},
one is allowed, besides, to state that each product $I_{j}CI_{j}$
is linearly dependent on the projector $I_{j}$.

It is interesting to observe that the relation \eqref{eq:Esp_Proiezione_j}
is characteristic of primitive projectors. Let, in fact, $I$ be a
projector such that for each observable $C$ it results:

\begin{equation}
ICI=cI\label{eq:Propr_Caratt_Ind_El1}
\end{equation}
where $c$ is a suitable real constant. Assuming that $I$ is not
primitive, they must exist two orthogonal projectors, $J_{1}$ and
$J_{2}$, such that it results:

\begin{equation}
I=J_{1}+J_{2}\text{ .}\label{eq:Propr_Caratt_Ind_El2}
\end{equation}
By applying \eqref{eq:Propr_Caratt_Ind_El1} to both $J_{1}$ and
$J_{2}$, moreover, one has:

\[
IJ_{1}I=y_{1}I\qquad\mbox{and}\qquad IJ_{2}I=y_{2}I
\]
where $y_{1}$ and $y_{2}$ are two suitable real constants. By substituting
these relations in \eqref{eq:Propr_Caratt_Ind_El2}, finally we get:

\[
J_{1}=y_{1}I\qquad\mbox{and}\qquad J_{2}=y_{2}I
\]
that necessarily implies that one between $J_{1}$ and $J_{2}$ must
be the zero observable and, therefore, that the projector $I$ is
primitive.

Projections are easily extended also to the case of pseudo-observables.
In fact, if $P\in\mathbb{P}$ is a pseudo-observable and $I$ is a
primitive projector, by making use of \eqref{eq:Forma_Compl_PO},
one immediately obtains:
\begin{equation}
IPI=IP_{\mathrm{R}}I+i\,IP_{\mathrm{I}}I=(p_{\mathrm{R}}+ip_{\mathrm{I}})I=\varpi I\label{eq:Proiezione_PO}
\end{equation}
where $p_{\mathrm{R}}$ and $p_{\mathrm{I}}$ are suitable real constants
and $\varpi\ideq p_{\mathrm{R}}+i\,p_{\mathrm{I}}$ a complex constant.

We now define \emph{dyadic form relative to the pair of primitive
projectors $\left(I_{j},I_{k}\right)$ with core $P$} the product:
$I_{j}PI_{k}$, where $P$ is a generic pseudo-observable. In next
subsection, we will be able to construct a useful basis for the space
$\mathbb{P}$ of the pseudo-observables formed by dyadic forms with
suitably chosen cores.

To this aim, it will be assumed that for each pair of primitive projectors
$\left(I_{j},I_{k}\right)$ there is always a pseudo-observable $P_{jk}$
such that the dyadic form relative to the pair of primitive projectors
with core $P_{jk}$ is nonzero.

\subsection{Dyad bases\label{subsec:Diadi}}

We will now prove that it is possible to construct a set of dyadic
form with suitably chosen pseudo-observable cores such that it constitutes
a basis for the whole space $\mathbb{P}$of the pseudo-observables.
Such dyadic forms will be called \textbf{\emph{dyads}}\footnote{\emph{Dyads} correspond in the Dirac formulation of quantum mechanics
to operators of the form: $\ket j\bra k$, where the indexes $j$
and $k$ denote distinct eigenvectors normalized to one, of a given
linear Hermitian operator. The name of \emph{dyad} will find justification
in subsection ``State vectors'' of the second paper.}, here denoted by the symbols $\Gamma_{jk}$, and satisfy the following
conditions:
\begin{enumerate}
\item $\Gamma_{jj}=I_{j}$
\item $\Gamma_{kj}=\Gamma_{jk}^{\dagger}$
\item $\Gamma_{jl}\Gamma_{l'k}=\delta_{l,l'}\,\Gamma_{jk}$
\end{enumerate}
where the Kronecker symbol $\delta_{l,l'}$ is, as usual, equal to
$1$ if the subscript indexes are equal and $0$ otherwise. In order
to do this, called $C_{jk}$ the core of the dyad $\Gamma_{jk}$,
one can observe that, by transposition, it results: 
\[
I_{k}C_{jk}I_{j}\neq0\;\Leftrightarrow\;I_{j}C_{jk}^{\dagger}I_{k}\neq0
\]
so that, for the second condition, one can put:
\[
C_{kj}=C_{jk}^{\dagger}\text{ .}
\]
For what regards the third condition, fixed a value $k_{0}$ of the
second index, for each index $j\neq k_{0}$ one can choose a pseudo-observable
$A_{j}$ such that it results: $\Phi_{jk_{0}}\ideq I_{j}A_{j}I_{k_{0}}\neq0$.
Observe now that, by the fifth property of the transposition, it results:
$\Phi_{jk_{0}}\Phi_{jk_{0}}^{\dagger}\geq0$ and therefore, by virtue
of \eqref{eq:Esp_Proiezione_j}, one has:

\[
\Phi_{jk_{0}}\Phi_{jk_{0}}^{\dagger}=I_{j}A_{j}I_{k_{0}}A_{j}^{\dagger}I_{j}=a_{j}^{2}I_{j}
\]
where $a_{j}$ is a suitable real positive constant. If we now put:
\[
C_{jk_{0}}=a_{j}^{-1}A_{j}
\]
One has:
\begin{equation}
\Gamma_{jk_{0}}\Gamma_{k_{0}j}=\Gamma_{jk_{0}}\Gamma_{jk_{0}}^{\dagger}=a_{j}^{-2}\Phi_{jk_{0}}\Phi_{jk_{0}}^{\dagger}=I_{j}=\Gamma_{jj}\label{eq:Gjk0_Gk0j}
\end{equation}
and, therefore, the first condition is satisfied for the index $j$.
Observe, now, that, by virtue of \eqref{eq:Esp_Proiezione_j}, one
has:

\begin{equation}
\Gamma_{k_{0}j}\Gamma_{jk_{0}}=\Gamma_{jk_{0}}^{\dagger}\Gamma_{jk_{0}}=I_{k_{0}}C_{jk_{0}}^{\dagger}I_{j}C_{jk_{0}}I_{k_{0}}=g_{j}I_{k_{0}}\label{eq:Gk0j_Gjk0-1}
\end{equation}
where $g_{j}$ is a suitable real positive constant. By the associative
property of the product, besides, it results:
\[
\left(\Gamma_{k_{0}j}\Gamma_{jk_{0}}\right)\Gamma_{k_{0}j}=\Gamma_{k_{0}j}\left(\Gamma_{jk_{0}}\Gamma_{k_{0}j}\right)
\]
that, by substituting relations \eqref{eq:Gjk0_Gk0j} and \eqref{eq:Gk0j_Gjk0-1},
allows one to obtain:
\[
g_{j}I_{k_{0}}\Gamma_{k_{0}j}=\Gamma_{k_{0}j}I_{j}\;\Rightarrow\;g_{j}=1
\]
by which it follows:
\begin{equation}
\Gamma_{k_{0}j}\Gamma_{jk_{0}}=I_{k_{0}}=\Gamma_{k_{0}k_{0}}\text{ .}\label{eq:Gk0j_Gjk0}
\end{equation}
Let's put, now, for $j\neq k$, $j\neq k_{0}$ and $k\neq k_{0}$:
\begin{equation}
C_{jk}=C_{jk_{0}}I_{k_{0}}C_{kk_{0}}^{\dagger}\label{eq:Def_Cjk}
\end{equation}
that immediately implies:
\begin{equation}
\Gamma_{jk}=\Gamma_{jk_{0}}\Gamma_{k_{0}k}\text{ .}\label{eq:Gjk0_Gk0k}
\end{equation}
The relation \eqref{eq:Gjk0_Gk0k} applies without any restriction
over the indexes $j$ and $k$. For $j=k$, in fact, the relation
reduces itself to \eqref{eq:Gjk0_Gk0j}; whereas for $j=k_{0}$ and
$k=k_{0}$, it immediately follows from the the fact that it is $\Gamma_{k_{0}k_{0}}=I_{k_{0}}$.
One should, besides, note that the definition \eqref{eq:Def_Cjk}
is well-posed, since, according to \eqref{eq:Gjk0_Gk0j} and \eqref{eq:Gk0j_Gjk0},
it results: 

\begin{eqnarray*}
\left(\Gamma_{jk_{0}}\Gamma_{k_{0}k}\right)\left(\Gamma_{jk_{0}}\Gamma_{k_{0}k}\right)^{\dagger} & = & \Gamma_{jk_{0}}\Gamma_{k_{0}k}\Gamma_{kk_{0}}\Gamma_{k_{0}j}=\Gamma_{jk_{0}}I_{k_{0}}\Gamma_{k_{0}j}=\\
 & = & \Gamma_{jk_{0}}\Gamma_{k_{0}j}=I_{j}\neq0
\end{eqnarray*}
and therefore, by the sixth property of the transposition, $\Gamma_{jk_{0}}\Gamma_{k_{0}k}\neq0$.
The dyadic forms with core given by \eqref{eq:Def_Cjk}, finally,
verify also the third condition of the dyad bases. In fact, one has:
\[
\Gamma_{jl}\Gamma_{lk}=\Gamma_{jk_{0}}\Gamma_{k_{0}l}\Gamma_{lk_{0}}\Gamma_{k_{0}k}=\Gamma_{jk_{0}}I_{k_{0}}\Gamma_{k_{0}k}=\Gamma_{jk_{0}}\Gamma_{k_{0}k}=\Gamma_{jk}
\]
and the property immediately follows if one consider also the orthogonality
relation between the projectors of the basis.

We can now prove an important property of the dyadic forms. For each
pseudo-observable $P$, let's consider the dyadic form relative to
the pair of primitive projectors $\left(I_{j},I_{k}\right)$ with
core $P$. By making use of \eqref{eq:Proiezione_PO} and of the properties
of the dyads, one has:
\begin{equation}
I_{j}PI_{k}=I_{j}P\,\Gamma_{kj}\Gamma_{jk}=\left(I_{j}\left(P\,\Gamma_{kj}\right)I_{j}\right)\Gamma_{jk}=(\varpi_{jk}\,I_{j})\Gamma_{jk}=\varpi_{jk}\Gamma_{jk}\label{eq:Componente_Diadica}
\end{equation}
that generalizes the projections.

By virtue of the postulate on the dyadic forms and we can now state
that the pseudo-observable space $\mathbb{P}$ is spanned by the set
of dyads $\left\{ \Gamma_{jk}\right\} $. If, in fact, $P$ is a generic
pseudo-observable, by making use of the decomposition \eqref{eq:Generatori_P}
and of \eqref{eq:Componente_Diadica}, it results:
\begin{equation}
P=\sum_{j,k}I_{j}PI_{k}=\sum_{j,k}\varpi_{jk}\,\Gamma_{jk}\label{eq:Scomp_PO}
\end{equation}
where the coefficients $\varpi_{jk}$, the \emph{dyadic components},
are suitable complex constants.

Dyads are, also, linearly independent. If, in fact, we consider a
linear combination of them, by means of a set of complex constants
$\left\{ \alpha_{jk}\right\} $, and set it to be equal to the zero
observable:

\begin{equation}
\sum_{j,k}\alpha_{jk}\,\Gamma_{jk}=0\label{eq:Rel_Lin_Ind_Diadi}
\end{equation}
after chosen any two indexes, $j'$ and $k'$, and multiplying the
relation \eqref{eq:Rel_Lin_Ind_Diadi} on the left by $I_{j'}$ and
on the right by $I_{k'}$, according to \eqref{eq:Ind_Alternativi},
one has:

\[
\alpha_{j'k'}\,\Gamma_{j'k'}=0\;\Rightarrow\;\alpha_{j'k'}=0
\]
from which follows the linear independence of the dyads.

The set of dyads $\left\{ \Gamma_{jk}\right\} $ therefore constitutes
a basis, that, in our assumptions, is finite or countable to the limit,
for the pseudo-observable space. This basis is equipotent to the Cartesian
product of the basis of primitive projectors by itself. Since the
bases of a vector space are mutually equipotent, an important consequence
of this result is that also the bases of primitive projectors associated
with the various spaces of compatible observables must be each other
equipotent.

Since the set of dyads $\left\{ \Gamma_{jk}\right\} $ constitutes
a basis for the space $\mathbb{P}$, each coefficient $\varpi_{jk}$
of the decomposition\eqref{eq:Scomp_PO} of a generic pseudo-observable
$P$ is, therefore, uniquely determined and will be called \emph{component
of the pseudo-observable according to the dyad $\Gamma_{jk}$}.

The dyad bases are therefore t natural extensions to the whole space
of pseudo-observables of the bases of pairwise orthogonal primitive
projectors and allow one to make explicit, in terms of components,
the operations of sum and of product of pseudo-observables, completing,
as already stated, their definition.

The components $\varpi_{jk}$ may be thought, in the finite-dimensional
cases, as elements of a matrix that will be called \emph{matrix associated
to the pseudo-observable with respect to the dyad basis}. On the basis
of this interpretation, we will call \emph{diagonal components} the
dyadic components of the form $\varpi_{jj}$. 

The concept of matrix associated will be used also in infinite-dimensional
spaces, that would be treated as limiting cases of finite-dimensional
ones.

Observe, now, that for a pseudo-observable $P$ it results $P^{\dagger}=P$
if and only if:

\begin{equation}
\varpi_{kj}=\varpi_{jk}^{*}\text{ .}\label{eq:Componenti_Osservabile}
\end{equation}
The matrices associated to such pseudo-observables are, therefore,
\emph{Hermitian}. We will, accordingly, call \emph{Hermitian pseudo-observables}
those satisfying the condition $P^{\dagger}=P$. The second property
of the transposition, section \ref{subsec:Trasposizione}, can so
be rephrased by stating that \emph{the observables are all and only
the Hermitian pseudo-observables}.

If $Q$ is another pseudo-observable, decomposed according the dyad
basis as follows:

\[
Q=\sum_{j,k}\theta_{jk}\,\Gamma_{jk}
\]
where $\theta_{jk}$ are suitable complex constants, one has:
\begin{equation}
P+Q=\sum_{j,k}\left(\varpi_{jk}+\theta_{jk}\right)\Gamma_{jk}\label{eq:Somma_PO_Comp}
\end{equation}
and
\begin{equation}
PQ=\sum_{j,k}\left(\sum_{l}\varpi_{jl}\,\theta_{lk}\right)\Gamma_{jk}\label{eq:Prod_PO_Comp}
\end{equation}
that allow one to calculate the sum and the product of two pseudo-observables,
being known the components of the pseudo-observables involved according
a given dyad basis.

It is to be observed that, according to \eqref{eq:Somma_PO_Comp}
and \eqref{eq:Prod_PO_Comp}, the matrix associated to the sum of
pseudo-observables is equal to the sum of the matrices associated
to the addends, while the matrix associated to the product of pseudo-observables
is equal to the matrix product of the matrices associated to the factors.
Fixed a dyad basis, there is, therefore, an isomorphism between the
ring of pseudo-observables and appropriate matrix rings. This isomorphism
makes, in particular, observables correspond to Hermitian matrices.
In this manner it is recovered and clarified the link, characteristic
of the Dirac formalism of quantum mechanics, between these two entities.

It will be, now, clarified better the tie between dyad basis and basis
of pairwise orthogonal primitive projectors. We will say, first of
all, that a dyad basis $\left\{ \Gamma_{jk}\right\} $ is \emph{associated
}to the basis $\left\{ I_{j}\right\} $ of\emph{ }pairwise orthogonal
primitive projectors, if all the dyadic forms of the first are relative
to pairs of projectors of the basis $\left\{ I_{j}\right\} $. This
stated, let $\left\{ \Gamma_{jk}\right\} $ and $\left\{ \Gamma'_{jk}\right\} $
be two dyad bases associated to the same basis $\left\{ I_{j}\right\} $
of pairwise orthogonal primitive projectors. By virtue of the \eqref{eq:Scomp_PO},
there must exist complex constants $\gamma_{jk}$ such that it results:

\begin{equation}
\Gamma'_{jk}=\gamma_{jk}\,\Gamma_{jk}\text{ .}\label{eq:Rel_Basi_Diadi_Eq}
\end{equation}
The second property of dyads, besides, implies:
\begin{equation}
\gamma_{jk}^{*}=\gamma_{kj}\text{ .}\label{eq:Simm_Coef_Diadi_Eq}
\end{equation}
By virtue of this relation and of the second and third property of
the dyads, one, therefore, has:
\[
I_{j}=\Gamma'_{jk}\left(\Gamma'_{jk}\right)^{\dagger}=\left|\gamma_{jk}\right|^{2}I_{j}\;\Rightarrow\;\left|\gamma_{jk}\right|^{2}=1
\]
where $\left|\gamma_{jk}\right|^{2}\ideq\gamma_{jk}^{*}\gamma_{jk}$
is the square of the \emph{modulus} of the complex constant $\gamma_{jk}$.
This last relation implies that the coefficients $\gamma_{jk}$ are
\emph{phase factors}, i.e. complex constants of the form: 
\begin{equation}
\gamma_{jk}=e^{i\vartheta_{jk}}\label{eq:Fattori_Fase_Diadi_Eq}
\end{equation}
in which the real constants $\vartheta_{jk}$, by virtue of \eqref{eq:Simm_Coef_Diadi_Eq},
satisfy the relation:
\begin{equation}
\vartheta_{kj}=-\vartheta_{jk}\text{ .}\label{eq:Simm_Fasi_Diadi_Eq}
\end{equation}
Having arbitrarily chosen an index $k_{0}$ and put:
\begin{equation}
\vartheta_{j}\ideq\vartheta_{jk_{0}}\label{eq:Fasi_Diadi_Eq}
\end{equation}
by the second and third property of the dyads and the relations \eqref{eq:Rel_Basi_Diadi_Eq},
\eqref{eq:Fattori_Fase_Diadi_Eq} and \eqref{eq:Fasi_Diadi_Eq}, one
has:
\begin{equation}
\Gamma'_{jk}=\Gamma'_{jk_{0}}\left(\Gamma'_{kk_{0}}\right)^{\dagger}=e^{i\left(\vartheta_{j}-\vartheta_{k}\right)}\Gamma_{jk}\label{eq:Basi_Diadi_Eq}
\end{equation}
that compared to \eqref{eq:Rel_Basi_Diadi_Eq}, by means of \eqref{eq:Fattori_Fase_Diadi_Eq},
allows one to set:
\begin{equation}
\vartheta_{jk}=\vartheta_{j}-\vartheta_{k}\label{eq:Diff_Fasi_Diadi_Eq}
\end{equation}
where the \emph{phases} $\vartheta_{j}$ are arbitrary real constants.
In general we will call \emph{equivalent} two dyad bases associated
to the same basis of primitive projectors.

\subsection{Change of dyad basis\label{subsec:Cambiamento-di-base}}

We will now show as to do a change of dyad basis, i.e. as to express
the dyads of a given basis $\left\{ \Gamma_{jk}\right\} $, associated
to the basis $\left\{ I_{j}\right\} $ of pairwise orthogonal primitive
projectors, in terms of the dyads of another basis $\left\{ \Gamma'_{j'k'}\right\} $,
associated to the basis $\left\{ I'_{j'}\right\} $ of pairwise orthogonal
primitive projectors. At this aim, one has to start to choose an index
$k_{0}$. Since it results:
\[
I_{k_{0}}=\sum_{k'}I_{k_{0}}I'_{k'}\neq0
\]
one may choose an index $k'_{0}$ such that it is $I_{k_{0}}I'_{k'_{0}}\neq0$.
The pseudo-observable $I_{k_{0}}I'_{k'_{0}}$, besides, may be expressed,
according the \eqref{eq:Scomp_PO}, as a linear combination of the
dyads of the basis $\left\{ \Gamma_{jk}\right\} $. By virtue of the
\eqref{eq:Componente_Diadica} one finds that it results:
\[
I_{k_{0}}I'_{k'_{0}}=\sum_{l}\alpha_{l}\,\Gamma_{k_{0}l}
\]
where the $\alpha_{l}$ are suitable non-null complex constants. By
this relation and the third property of the dyads, one obtains:
\[
\Gamma_{jk_{o}}I'_{k'_{0}}=\Gamma_{jk_{o}}\left(I_{k_{0}}I'_{k'_{0}}\right)=\sum_{l}\alpha_{l}\,\Gamma_{jl}\text{ .}
\]
Note, also, that it results:
\[
\left(\Gamma_{jk_{o}}I'_{k'_{0}}\right)\left(\Gamma_{kk_{o}}I'_{k'_{0}}\right)^{\dagger}=\sum_{l,l'}\alpha_{l}\alpha_{l'}^{*}\,\Gamma_{jl}\Gamma_{l'k}=\sum_{l}\left|\alpha_{l}\right|^{2}\Gamma_{jk}
\]
by which, after putting $\sum_{l}\left|\alpha_{l}\right|^{2}\ideq1/g^{2}$,
one obtains:
\begin{equation}
\Gamma_{jk}=g^{2}\left(\Gamma_{jk_{o}}I'_{k'_{0}}\right)\left(\Gamma_{kk_{o}}I'_{k'_{0}}\right)^{\dagger}\text{ .}\label{eq:Scomp_Diadi_I}
\end{equation}
Let we, now, decompose the products $\Gamma_{jk_{o}}I'_{k'_{0}}$
according the dyads of the basis $\left\{ \Gamma'_{j'k'}\right\} $:
\begin{equation}
\Gamma_{jk_{o}}I'_{k'_{0}}=\sum_{l}\beta_{lj}\,\Gamma'_{lk'_{0}}\label{eq:Scomp_Diadi_II}
\end{equation}
where $\beta_{lj}$ are suitable complex constants and having used
\eqref{eq:Componente_Diadica} and the relationships among the projectors
of a same basis. By substituting \eqref{eq:Scomp_Diadi_II} in \eqref{eq:Scomp_Diadi_I},
therefore, one has:
\begin{equation}
\Gamma_{jk}=\sum_{l,l'}\left(g\,\beta_{lj}\right)\left(g\,\beta_{l'k}\right)^{*}\Gamma'_{lk'_{0}}\Gamma'_{k'_{0}l'}=\sum_{l,l'}\omega_{lj}\omega_{l'k}^{*}\,\Gamma'_{ll'}\label{eq:Cambiamento_Base_Diadi_I}
\end{equation}
having set: $\omega_{lj}\ideq g\,\beta_{lj}$. Equation \eqref{eq:Cambiamento_Base_Diadi_I},
expressing the relation among the dyads of a basis and those of another
basis, may be put in a form of particular interest by introducing
the \emph{pseudo-observable $\Omega$}\textit{\emph{ }}\emph{of the
change from the basis }\textit{\emph{$\left\{ \Gamma_{jk}\right\} $}}\emph{
to the basis }\textbf{\textit{\emph{$\left\{ \Gamma'_{jk}\right\} $}}}:
\begin{equation}
\Omega\ideq\sum_{l,m}\omega_{lm}\,\Gamma'_{lm}\text{ .}\label{eq:PO_Camb_Base}
\end{equation}
It is easily verified that:
\begin{equation}
\Gamma_{jk}=\Omega\Gamma'_{jk}\Omega^{\dagger}\text{ .}\label{eq:Cambiamento_Basi_Diadi_II}
\end{equation}
Due to the symmetry between the two basis, the relation \eqref{eq:Cambiamento_Basi_Diadi_II}
must be invertible, and this implies that the pseudo-observable $\Omega$
admits left inverse. By putting $k=j$ in \eqref{eq:Cambiamento_Basi_Diadi_II}
and remembering the first property of the dyads, it results:
\begin{equation}
I_{j}=\Gamma_{jj}=\Omega\Gamma'_{jj}\Omega^{\dagger}=\Omega I'_{j}\Omega^{\dagger}\label{eq:Cambiamento_Basi_Indicat}
\end{equation}
by summation over the index $j$ and recalling the closure relation
\eqref{eq:Rel_Chiusura}, one obtains:
\[
\Omega\Omega^{\dagger}=1\text{ .}
\]
The change of basis pseudo-observable so admits right inverse too
and it is therefore \emph{invertible}. Its \emph{inverse}, besides,
due to the previous relation, coincides with its transposition:
\begin{equation}
\Omega^{-1}=\Omega^{\dagger}\;\Rightarrow\;\Omega^{\dagger}\Omega=\Omega\Omega^{\dagger}=1\text{ .}\label{eq:PO_Unitaria}
\end{equation}
The relations \eqref{eq:PO_Unitaria}, expressed in terms of components,
are equivalent, by virtue of \eqref{eq:PO_Camb_Base} and \eqref{eq:Prod_PO_Comp},
to the following:
\begin{equation}
\sum_{l}\omega_{lj}^{*}\omega_{lk}=\sum_{l}\omega_{jl}\omega_{kl}^{*}=\delta_{jk}\text{ .}\label{eq:Rel_Unitarieta}
\end{equation}
The matrix associated to $\Omega$ is, therefore, \emph{unitary}.
We will then call \emph{unitary} every pseudo-observable satisfying
\eqref{eq:PO_Unitaria}.

Some useful properties, of immediate verification, of unitary pseudo-observables
are the following:
\begin{enumerate}
\item \textbf{If $\Omega$ is an unitary pseudo-observable, its transposition,
$\Omega^{\dagger}$, and its inverse, $\Omega^{-1}$, which, moreover,
coincide with each other, are unitary too.}
\item \textbf{If $\Omega_{1}$ and $\Omega_{2}$ are two unitary pseudo-observables,
their product, $\Omega_{1}\Omega_{2}$, is an unitary pseudo-observable
also.}
\end{enumerate}
As first important example of change of basis, we will regard that
induced by the pseudo-observable:
\begin{equation}
S_{j_{0}j_{1}}\ideq1-I_{j_{0}}-I_{j_{1}}+\Gamma_{j_{0}j_{1}}+\Gamma_{j_{1}j_{0}}\text{ .}\label{eq:PO_Scambio}
\end{equation}
It is immediately verified that $S_{j_{0}j_{1}}$ is Hermitian and
unitary. The change of basis associated to it exchanges in the dyads
the projector $I_{j_{0}}$ with $I_{j_{1}}$.

Another interesting change of basis is that associated to the pseudo-observable:
\begin{equation}
\Omega_{\vartheta}\ideq\sum_{j}e^{i\vartheta_{j}}\,I_{j}\label{eq:Camb_Basi_Equiv}
\end{equation}
where the $\vartheta_{j}$ are real constants. It is immediate to
verify that $\Omega_{\vartheta}$ is unitary and that the relative
change of basis corresponds to one that makes it to pass from a dyad
basis to an \emph{equivalent} one, according to what seen at the end
of subsection \ref{subsec:Diadi}.

In such a manner the matrix structure of Heisenberg's formulation
of quantum mechanics is fully recovered and justified.

\section{Conclusions}

\subsection{Summary of Postulates\label{subsec:Riepilogo-dei-postulati}}

At this point it is necessary to make a summary of the postulates
introduced, commenting them appropriately to better understand their
meaning:
\begin{enumerate}
\item Each physical system is characterized by a set of measurable properties,
called \emph{observables}. An observable is characterized by all the
possible outcomes of its measurement (the \emph{observable spectrum}).
\item Two observables are said to be \emph{compatible} if you can measure
them simultaneously and independently, in the sense that measurement
of one does not affect the measurement outcomes of the other. \textbf{There
do exist incompatible observables}.
\item For two given compatible observables it is possible to define their
sum (if they are homogeneous among themselves) and their product in
terms of the their spectra. For every observable it is possible to
determine at least a set of observables, of which it is a member,
that is \emph{complete}, i.e. such that each observable compatible
with all the observables of the set may be expressed as a function
of them. The complete set of compatible observables admits an unique
basis of pairwise orthogonal primitive projectors, such that each
observable compatible with the observables of this set may be expressed
as linear combinations of the projectors of the basis.
\item It is \textbf{not possible} to give a definition of the sum or the
product of two incompatible observables based on physical based arguments.
One can, however, consider the space $\mathbb{P}$, formed by the
\textbf{formal expressions} obtained by sum and product of observables
(compatible or not). Such expressions, which have physical meaning
only when they involve solely observables compatible among themselves,
are called \emph{pseudo-observable}s. The \emph{space of the pseudo-observables}
is an algebraic structure whose construction is bound only by the
condition of being consistent with the properties of the algebra of
compatible observables. From a physical point of view it is appropriate
to follow a criterion of \emph{maximum significance}, in the spirit
of the \emph{Occam razor principle}, according to which hypotheses
will be taken to utmost reduce the entities without immediate physical
interpretation. To this end, it is assumed that the space has a \emph{unitary
ring} structure with respect to sum and product operations. The existence
of incompatible observables implies that the product is generally
not commutative.
\item In accordance with the criterion of maximum significance, a \emph{transposition
operator} ($\dagger$) is introduced, which reverses the observation
order of a set of observables, characterized by the six properties
(assumed by hypothesis) presented and commented at the beginning of
the subsection \ref{subsec:Trasposizione}.
\item It is introduced the pseudo-observable $i$, formally defined as a
pseudo-observable that commutes with every other one and such that
it results: $i^{\dagger}=-i$ and $i^{2}=-1$. By means of the pseudo-observable
$i$ it is possible write down the pseudo-observables in \emph{complex
form }and so put them in biunivocal correspondence with pairs of observable
(\emph{real} and \emph{imaginary part}).
\item It is, last, assumed that, for each pair of primitive projectors of
a basis, there is a nonzero associated dyadic form. This postulate
allows to extend the bases of pairwise orthogonal primitive projectors,
that span complete subspaces of compatible observables, to \emph{dyad
bases,} that span the whole space $\mathbb{P}$ of the pseudo-observables.
The decomposition of the pseudo-observables according the dyads of
a basis allows, finally, one to write explicit expressions for sums
and products of observables, compatible or not, or pseudo-observables.
\end{enumerate}

\subsection{Observers and meta-observers}

In this paper it was shown how the algebraic structure of quantum
mechanics is the \textbf{unique} one by means of which it is possible
to give a coherent description of observing a set of properties, not
all compatible with each other, and that introduces the minimum possible
number of non-observable entities.\emph{ The incomprehensible oddness
of quantum mechanics thus becomes a necessary characteristic of the
logic structure of a ``reality'' understood as a coherent construction
derived from a set of observations, or as a result of a dialectical
process of exchange of information between all the observers}.

The measurement problem, transformations and time evolution in the
framework of the algebra of the pseudo-observables will be treated
in the second paper, where it will be also revealed the full structure
of C{*}-algebra of this new algebra, and in the third. 

In conclusion of this paper, we want to discuss about the \textbf{immutability},
in space and time, of the laws of physics.

Even if the problem of the eternal validity of physical laws ultimately
refers to the historicizing of the evolution of the Universe and makes
obvious reference to the historical becoming of things; it is also
true that there is no History without Observers and without Observers
able to communicate with each other through the space and the time.
But communication aggregates the Observers, generating others of higher
level (\emph{meta-observers}). Thus the whole Humanity, in its historical
becoming, acquires the status of a meta-observer: the result of the
observational process of such a meta-observer is what we call (our)
``physical reality''.

An important point to clarify, is that the aggregation of the observers
into higher-level ones does not imply that the lasts sum up the perspective
of their ``parts''. As Anderson well stated in 1972\citep{Anderson1972},
``More Is Different'', and so it happens for the perspective of
higher-level observer. To make a simple example, it is well known
that our mind arises ``in some manner'' by the neuron activity.
Each neuron receive inputs and send output, and there is no doubt
that may be consider a simple observer. Our mind is the result of
the exchange of information of our neurons, and so is an observer
of higher level with respect to them, but cannot access the perspective
of any single neuron: \textbf{its perspective is different}!

Even inanimate matter can act as a ``passive'' observer, bringing
in its structure the traces of its own becoming. In the act in which
a researcher reads and interprets those traces, an one-way communication
is established, which still give rise to a meta-observer, of higher
level respect to its components.

If we therefore consider all possible interactions between observers,
active or passive, throughout the course of historical becoming, we
finally find a single entity that we will call the \emph{universal
observer} (or \emph{super-observer}), of level higher than any other
observer, necessarily \emph{atemporal} and \emph{delocalized}.

In conclusion:
\begin{enumerate}
\item What we call ``reality'' is an emergent property that arises by
the internal communications of a closed network of observers.
\item The logical structure of a such ``reality'' reflects the inner logical
structure of the observers, and this explains ``The Unreasonable
Effectiveness of Mathematics in the Natural Sciences'' quoted by
Wigner in 1960\citep{Wigner1960}
\item The hierarchical structure of the observers, aggregated in higher
and higher meta-levels till to that of the universal (meta-)observer,
justifies the validity of the laws of physics in \textbf{every} place
of the Universe and at \textbf{all} times. 
\end{enumerate}

\ack{}{}

I want to kindly thank David Edwards for a useful discussion that
gave me the opportunity to better clarify the role of the observers
in the theory and their hierarchical relationships.

\bibliographystyle{unsrt}
\addcontentsline{toc}{section}{\refname}\bibliography{../Bibliography}

\end{document}